\documentclass[a4paper,11pt]{article}
\usepackage[english]{babel}
\usepackage[utf8]{inputenc}
\usepackage[T1]{fontenc}
\usepackage{lmodern}
\usepackage{mathptmx}%
\usepackage[scaled=.95]{helvet} %
\usepackage[left=30mm,right=30mm,top=30mm,bottom=30mm,marginparwidth=17.5mm]{geometry}
\usepackage[hyphens]{url}
\usepackage{hyperref}
\usepackage{doi}
\usepackage{graphics}
\usepackage{graphicx}
\usepackage{color}
\usepackage{microtype}
\usepackage[table,xcdraw]{xcolor}
\usepackage[round]{natbib}
\usepackage{tikz}
\usepackage{amsmath}
\usepackage{amsfonts}
\usepackage{amsthm}
\usepackage{amssymb}
\usepackage{amsbsy}
\usepackage{dsfont}
\usepackage{mathtools}
\usepackage{bm}
\usepackage{bbm}
\usepackage[font=small]{caption}
\usepackage{subcaption}
\usepackage{float}
\usepackage{rotating}
\usepackage{stfloats} %
\usepackage{flafter} %
\usepackage{titling}
\usepackage[auth-lg]{authblk} %
\usepackage{fancyhdr} %
\usepackage{appendix}
\usepackage{booktabs}
\usepackage{enumitem}
\usepackage{nicefrac}
\usepackage[capitalise,nameinlink,sort&compress]{cleveref} %

\usepackage{siunitx}

\usepackage{numprint}%
\npdecimalsign{.}
\npthousandsep{,}

\usetikzlibrary{arrows,shapes,trees,shapes.geometric}

\hypersetup{
    colorlinks,
    breaklinks,
    linkcolor={blue!75!black},
    citecolor={blue!75!black},
    urlcolor={blue!75!black},
    pdftitle={Validating Deep-Learning Weather Forecast Models on Recent High-Impact Extreme Events},
    pdfauthor={Olivier C. Pasche, Jonathan Wider, Zhongwei Zhang, Jakob Zscheischler, and Sebastian Engelke},
    pdfkeywords={}
}

\theoremstyle{plain}

\theoremstyle{remark}

\newcommand{\indep}{\perp\kern-5.5pt\perp}

\newenvironment{significance}
  {\begin{abstract}}
  {\end{abstract}}

\let\LTXmaketitle\maketitle
\renewcommand{\maketitle}{{\bf\LTXmaketitle}} %

\title{Validating Deep-Learning Weather Forecast Models on Recent High-Impact Extreme Events}
\newcommand{\shorttitle}{Validating D.L.\ Weather Forecast Models on Recent High-Impact Extreme Events}

\author[a,*]{Olivier~C.~Pasche}
\author[b,c,*]{Jonathan~Wider}
\author[a,*]{Zhongwei~Zhang}
\author[b,c,d]{Jakob~Zscheischler}
\author[a]{Sebastian~Engelke}

\affil[a]{Research Institute for Statistics and Information Science, University of Geneva, Switzerland}
\affil[b]{Department of Compound Environmental Risks,\par Helmholtz~Centre~for~Environmental~Research~--~UFZ, Leipzig, Germany}
\affil[c]{Center for Scalable Data Analytics and Artificial Intelligence (ScaDS.AI),\par Dresden/Leipzig, Germany}
\affil[d]{Department of Hydro Sciences, TUD Dresden University of Technology, Dresden, Germany}
\affil[*]{Authors contributed equally}

\date{}

\begin{document}

\pagestyle{fancy}
\fancyhead{} %
\fancyhead[C]{\small\sc\shorttitle}
\renewcommand{\headrulewidth}{0pt} %

\maketitle

\begin{abstract}
The forecast accuracy of machine learning (ML) weather prediction models is improving rapidly, leading many to speak of a “second revolution in weather forecasting”. With numerous methods being developed and limited physical guarantees offered by ML models, there is a critical need for a comprehensive evaluation of these emerging techniques. While this need has been partly fulfilled by benchmark datasets, they provide little information on rare and impactful extreme events or on compound impact metrics, for which model accuracy might degrade due to misrepresented dependencies between variables. To address these issues, we compare ML weather prediction models (GraphCast, PanguWeather, and FourCastNet) and ECMWF's high-resolution forecast system (HRES) in three case studies: the 2021 Pacific Northwest heatwave, the 2023 South Asian humid heatwave, and the North American winter storm in 2021. We find that ML weather prediction models locally achieve similar accuracy to HRES on the record-shattering Pacific Northwest heatwave but underperform when aggregated over space and time. However, they forecast the compound winter storm substantially better. We also highlight structural differences in how the errors of HRES and the ML models build up to that event. The ML forecasts lack important variables for a detailed assessment of the health risks of the 2023 humid heatwave. Using a possible substitute variable, prediction errors show spatial patterns with the highest danger levels over Bangladesh being underestimated by the ML models. Generally, case-study-driven, impact-centric evaluation can complement existing research, increase public trust, and aid in developing reliable ML weather prediction models.
\end{abstract}

\begin{significance}
With the performance of machine-learning-based weather forecasting models improving rapidly, thorough analyses are needed to ensure that their forecasts are accurate and reliable before deploying them in operational settings.
Existing evaluations often reduce forecast performance to a few metrics, potentially obscuring rare but systematic errors. This is especially problematic for high-impact extreme events, which, by definition, are rare in the data but often substantially affect society. 
In a detailed analysis of three extreme events, we observe that, although machine learning (ML) models generally outperform the best physics-based numerical weather prediction (NWP) model on benchmark datasets, they do not consistently do so for the studied extreme events or compound impact metrics and lack some impact-relevant variables.
\end{significance}
    
\newpage

\section{Introduction}
\label{s:introduction}

In recent years, the performance of machine learning (ML) weather forecast models has improved drastically \citep{raspWeatherBenchBenchmarkNext2024}, leading some authors to speak of a ``rise'' of ML methods in weather forecasting \citep{benBouallegueRiseDataDrivenWeather2024}, or even a second revolution of the field. While some studies focused on short-term predictions \citep[``nowcasting'',][]{espeholtDeepLearningTwelve2022, leinonenLatentDiffusionModels2023, andrychowicz2023deep}, and long-term subseasonal-to-seasonal forecasting \citep[two weeks to two months ahead,][]{weynSubSeasonalForecasting2021,lopez-gomezGlobalExtremeHeat2023}, much work concentrated on the medium range \citep{raspDataDrivenMedium2021, pathakFourCastNetGlobalDatadriven2022,
nguyenClimaXFoundationModel2023,
chenFengWuPushingSkillful2023,
biAccurateMediumrangeGlobal2023,
kochkovNeuralGeneralCirculation2024, lamLearningSkillfulMediumrange2023, chenFuXiCascadeMachine2023, nguyenScalingTransformerNeural2023, priceGenCastDiffusionbasedEnsemble2023}, i.e., forecasting days to two weeks into the future. 

The established technique for weather forecasting in the medium range is numerical weather prediction (NWP), which is based on evolving an estimate of the current weather state constructed from observations through time under differential equations. Therefore, the point of comparison for ML approaches is ECMWF's Integrated Forecasting System \citep{owensECMWFForecastUser2018}, including its high-resolution forecast system (HRES), which is generally considered to be the most reliable NWP model for global deterministic weather forecasts. The latest ML weather models match or even outperform HRES in terms of overall summary scores across many variables, pressure levels, and prediction lead times \citep{raspWeatherBenchBenchmarkNext2024}. In addition to performance, other reasons to consider ML-based weather forecasting include energy efficiency during operations and improved inference speed. ML models that supplement or replace parts of the weather forecasting pipeline are increasingly seen as a realistic possibility \citep{bauerWhatIfNumerical2024, benBouallegueRiseDataDrivenWeather2024}. ECMWF is already publishing forecast data produced with its own AI model, AIFS \citep{langAIFSECMWFDatadriven2024}, as part of its experimental suite.

Given these recent advances and the immense importance of accurate and robust weather forecasting to many aspects of human life, thorough analyses are necessary before operationalizing ML weather prediction models. 
As extreme weather events often have severe impacts \citep{zscheischlerTypologyCompoundWeather2020, seneviratneWeatherClimateExtreme2021}, such as crop loss, wildfires, and floods, effective mitigation measures require accurate predictions in the tails of the distribution. 

While ML-based weather forecasts can achieve high overall accuracy, their performance for extreme events is not well understood. ML models generally face fundamental difficulties during extrapolation and generalization to unseen domains, and good test accuracy estimates do not guarantee good performance outside the range of previous observations, or in regions of the input space where observations were scarce \citep{ElemStatLearn, watsonMachineLearningApplications2022}.

Summary scores, like the root mean squared error (RMSE), play a central role in the evaluation of ML weather prediction models. Typically, one score is computed for each lead time, predicted variable, and (pressure) level to quantify the model's performance over the entire test set \citep[see, e.g., scorecards in][]{raspWeatherBenchBenchmarkNext2024}. Several other aspects of ML forecasts have also been studied in the literature. For instance, \citet{bonavitaLimitationsCurrentMachine2024}, \citet{lamLearningSkillfulMediumrange2023} and \citet{raspWeatherBenchBenchmarkNext2024} examined the smoothness of the predictions and found that most ML models tend to blur predictions for long lead times as a consequence of the way these models are conceptualized and trained. \cite{bonavitaLimitationsCurrentMachine2024} studied PanguWeather, one of the best-performing ML models, and found it to be worse at maintaining physical balances than ECMWF's HRES. 

\citet{gmd-17-2347-2024} summarized some of the extreme-event evaluations performed for the latest generation of ML weather forecast models. In previous work, extreme temperatures (both hot and cold) were studied by comparing threshold exceedances of predictions and ground truth data \citep{benBouallegueRiseDataDrivenWeather2024, lamLearningSkillfulMediumrange2023, olivettiDatadrivenModelsBeat2024}. Other types of investigated extreme events include tropical cyclones, atmospheric rivers, and storm systems \citep{magnussonExploringMachinelearningForecasts2023, benBouallegueRiseDataDrivenWeather2024, lamLearningSkillfulMediumrange2023, charlton-perezAIModelsProduce2023}. While some studies have looked into individual events, many types of extremes are still under-explored, especially on a case-study level.

In addition, little attention has been paid to impact metrics that combine multiple predicted variables, or to events where accurate assessment of their spatial or temporal extent is important. The compounding effect of multiple variables in space and time can lead to particularly large impacts \citep{zscheischlerTypologyCompoundWeather2020}. Examining prediction performance for these events in case studies is necessary to increase public trust in ML models, and also has the potential to uncover rare systematic errors in the ML model predictions that might be hidden by summary scores.

This study evaluates the ability of three popular ML weather prediction models to accurately forecast relevant impact metrics of extreme weather events through three case studies. 
The ML models GraphCast \citep{lamLearningSkillfulMediumrange2023}, PanguWeather \citep{biAccurateMediumrangeGlobal2023}, and FourCastNet \citep{pathakFourCastNetGlobalDatadriven2022} are compared to IFS HRES \citep{owensECMWFForecastUser2018} for the 2021 Pacific Northwest heatwave, the 2023 South Asian humid heatwave, and the 2021 North American winter storm.

\section{Data and Models}
\label{s:data-and-methods}

\subsection{Data}
\label{ss:data}
In this paper, we use two kinds of data: ERA5 reanalysis data \citep{hersbachERA5GlobalReanalysis2020} and ECMWF HRES analysis data.
All ML models considered in this study were trained on ERA5, which is produced using data assimilation, i.e., by combining observations with short-range forecasts to obtain a ``best guess'' of the actual weather state. ERA5 has a horizontal resolution of $\qty{0.25}{\degree} \times \qty{0.25}{\degree}$, an hourly temporal resolution, and provides estimates of many atmospheric, land, and oceanic climate variables over the globe from 1940 to present.
The ML models GraphCast and FourCastNet have an internal time step of \qty{6}{\hour} and were trained on a subset of ERA5 at 00:00, 06:00, 12:00, and 18:00 UTC. Therefore, we also restrict our analyses to these times of day. 

We use HRES forecasts of versions 47r1, 47r2, and 47r3 from ECMWF's Integrated Forecasting System. They have a horizontal resolution of $\qty{0.1}{\degree} \times \qty{0.1}{\degree}$, and we downsample them to the $\qty{0.25}{\degree} \times \qty{0.25}{\degree}$ grid using the default Meteorological Interpolation and Regridding (MIR) library in the ECMWF Meteorological Archival and Retrieval System (MARS). ERA5 and HRES data can be retrieved from online archives. HRES forecasts initialized at 00:00 UTC or 12:00 UTC are archived for lead times up to \qty{10}{\day}, while forecasts initialized at 06:00 UTC and 18:00 UTC are only available for lead times up to \qty{3.75}{\day}. 
To ensure a fair comparison, we use ``HRES forecast at step 0'' (HRES-fc0) as the ground truth for HRES forecasts. If ERA5 was used instead, HRES would have a non-zero error at lead time \qty{0}{h}.
We use HRES forecast data with lead times ranging from \qty{0}{\hour} to the maximum available length in steps of \qty{6}{\hour}, so that the lead times match those of the ML forecasts.

A difference between our comparison study and \cite{benBouallegueRiseDataDrivenWeather2024} is the data used for initializing and evaluating ML models.
In both cases, they used HRES data to ensure fairness in an operational context, since ERA5 reanalysis data are simply not available at the time of an operational prediction.
From a ML perspective, this will unfortunately bring disadvantages to ML models because they are trained on ERA5 data.
Here, we choose the conventional approach in ML studies \citep{pathakFourCastNetGlobalDatadriven2022,biAccurateMediumrangeGlobal2023,lamLearningSkillfulMediumrange2023,chenFuXiCascadeMachine2023} and use ERA5 data to initialize and evaluate ML models.

\subsection{Machine Learning Models for Weather Forecasting}
\label{ss:methods}
We focus on three recent ML models in this work: FourCastNet, PanguWeather, and GraphCast. \Cref{tab:ML_model_features} summarizes the main characteristics of these models. More details on the variables predicted by these models are presented in Section~\ref{sups:lists_variables} of the Supplementary Materials.

\begin{table*}[tb]
\centering
\caption{Summary of key features of recent ML-based weather forecasting models.}
\resizebox{\textwidth}{!}{%
\begin{tabular}{lccc}\toprule
   & FourCastNet & PanguWeather & GraphCast \\\midrule
Resolution & $0.25^\circ \times 0.25^\circ$ & $0.25^\circ \times 0.25^\circ$ & $0.25^\circ \times 0.25^\circ$ \\
Architecture & Fourier Neural Operator & Transformer & Graph Neural Network \\
\# surface variables & 6 & 4 & 5 \\
\# atmospheric variables & 5 at 4 pressure levels (pl) & 5 at 13 pl & 6 at 37 pl \\
Training and validation & 1979--2017 & 1979--2017, 2019 & 1979--2017 \\
\# parameters (millions) &  74.7 & 256 & 36.7 \\
Fundamental timestep (\unit{\hour}) &  6 & 1, 3, 6, 24  & 6 \\
\bottomrule
\end{tabular}%
}
\label{tab:ML_model_features}
\end{table*}

\cite{biAccurateMediumrangeGlobal2023} trained four models with different lead times (1, 3, 6, and 24 \unit{\hour}). These models are combined during inference to achieve the minimum number of model executions for a given forecast lead time (``hierarchical temporal aggregation strategy''), thereby minimizing error accumulation \citep{biAccurateMediumrangeGlobal2023}. For instance, to forecast the weather state in 36 hours, using the 24h-model once plus two iterations of the 6h-model gives a more accurate forecast than simply using the 1h-model iteratively for 36 times. Because we only consider lead times that are multiples of \qty{6}{h} in our study, we use sequences of \qty{6}{h}- and \qty{24}{h}-model calls to achieve the smallest possible forecast error for PanguWeather.

We also highlight that GraphCast requires the weather states at two consecutive time points as inputs to each forecast, while the other two models only need one.
Furthermore, as shown in \cref{tab:ML_model_features}, for each time step the dimensionality of the input data (especially the number of pressure levels) of GraphCast is substantially higher than that of PanguWeather and FourCastNet.
These two distinctions might advantage GraphCast in comparison to the other two ML models, since using additional covariate information in principle tends to increase the potential predictive accuracy.

Although forecast data from various ML models are available on \mbox{WeatherBench 2} \citep{raspWeatherBenchBenchmarkNext2024}, they do not cover all periods we investigate (e.g., GraphCast forecasts are only available for 2018 and 2020).
We thus produced additional forecast data for more recent events by running the ML models ourselves.
More precisely, we implemented the inference of the three ML models by directly leveraging their pre-trained models released on GitHub. Alternatively, the forecast data can be generated using the ECMWF library ``ai-models'', but note that the GraphCast model in the library is GraphCast operational (a smaller version than the one described in \citet{lamLearningSkillfulMediumrange2023}), which was pre-trained on ERA5 data from 1979 to 2017 and fine-tuned on HRES data from 2016 to 2021, and only includes atmospheric variables at 13 pressure levels as input and output.

\subsection{Initialization Times}
\label{ss:init-time}
ERA5 and HRES-fc0 differ in their assimilation windows \citep{lamLearningSkillfulMediumrange2023}. While observations up to three hours into the future are included in the assimilation for \mbox{HRES-fc0}, the lookahead for ERA5 varies between initialization times: three hours for forecasts initialized at 06:00/18:00 UTC and nine hours for forecasts initialized at 00:00/12:00 UTC.
To ensure an equal lookahead, \cite{lamLearningSkillfulMediumrange2023} compared ML-based and HRES forecasts initialized at 06:00/18:00 UTC for lead times up to the availability of HRES (\qty{3.75}{\day}).
Beyond this time limit, ML forecasts initialized at 06:00/18:00 UTC are compared with HRES forecasts initialized at the preceding 00:00/12:00 UTC.
We follow this mixed initialization methodology in one of our analyses in \cref{ss:pnw_heatwave}.

As shown in Section S.5.2 in the Supplementary materials of \cite{lamLearningSkillfulMediumrange2023}, the effect of unequal lookahead is small, particularly for long lead times. %
Therefore, for all analyses except the RMSE comparison in the first case study, we include all forecasts (00:00, 06:00, 12:00, and 18:00 UTC initialization times) in our analysis. Additionally, we extend the short HRES forecasts initialized at 06:00/18:00 UTC beyond lead times of \qty{4}{\day}; these forecasts are augmented with data from the forecasts initialized \qty{6}{\hour} prior to the 06:00/18:00 UTC initialization time, while increasing the lead time by \qty{6}{\hour}, so that the validity time of the forecast remains the same. This filling might disadvantage HRES, but it enables the analysis of a denser set of initialization and lead times.

\section{Case Studies}
\label{s:casestudies}

\subsection{2021 Pacific Northwest Heatwave}
\label{ss:pnw_heatwave}
\begin{figure*}[tb]%
    \centering%
    \includegraphics[width=\textwidth]{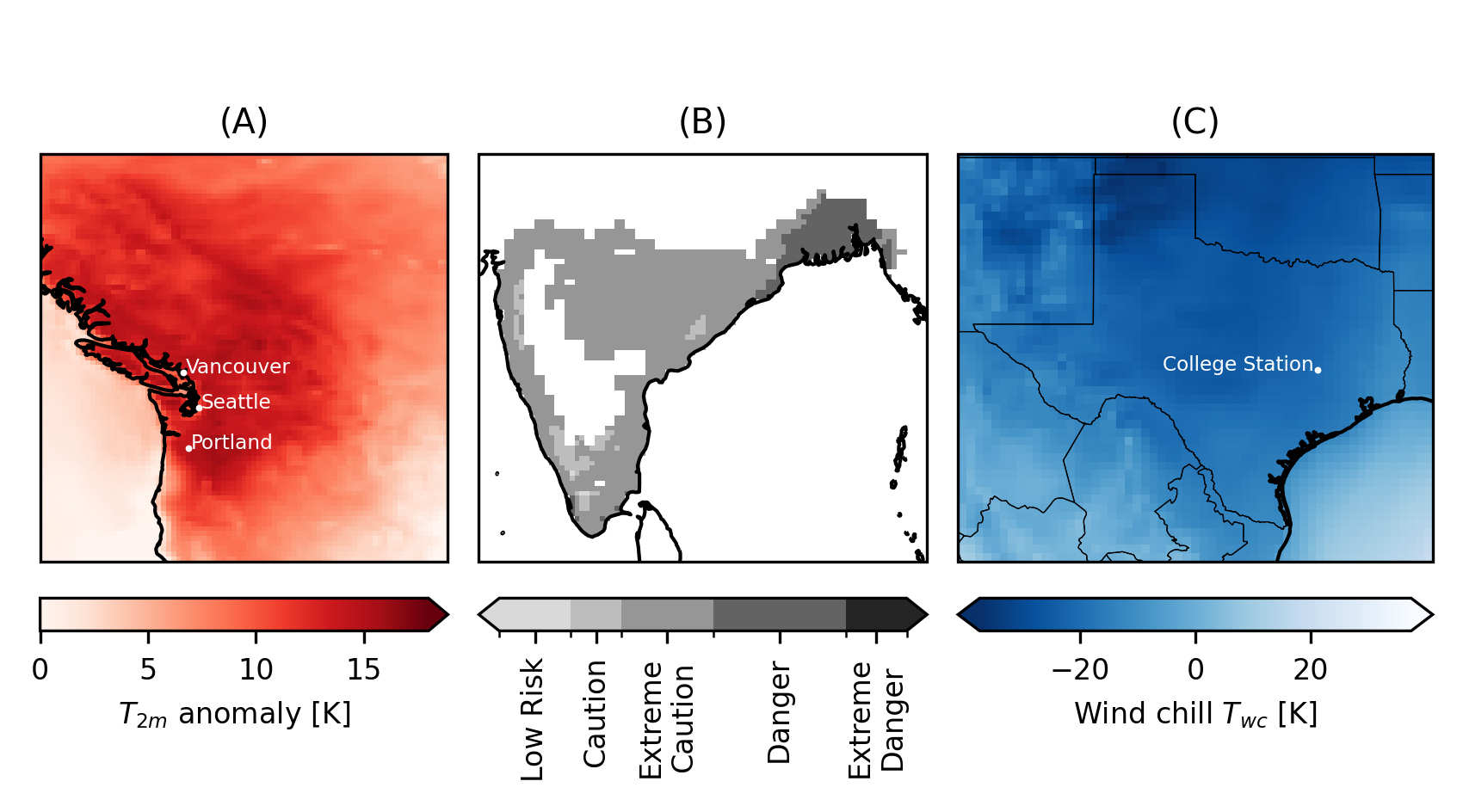}
    \caption[]{Magnitudes of the three events analyzed in this paper. (A) 2021 Pacific Northwest heatwave. Shown is the \SI{2}{\meter} temperature anomaly averaged over 27--29 June 2021, the peak of the heatwave. (B) 2023 South Asian humid heatwave. Shown is the category of maximum daily Heat Index $HI$, as defined in \cref{ss:shape_files}, averaged over 17--20 April 2023 in India and Bangladesh. (C) 2021 North American winter storm. Shown is the wind chill index $T_{wc}$, as defined in \cref{ss:na_winterstorm}, on 12:00 UTC, 15 February 2021.}
    \label{f:IntroPlots}
\end{figure*}

In this first case study, we investigate a record-shattering extreme temperature event. In late June 2021, a heatwave of unprecedented magnitude hit the Pacific Northwest with temperatures reaching up to \qty{49.6}{\celsius}, beating the all-time record for Canada by \qty{4.6}{\kelvin} (\cref{f:IntroPlots}A). 
Even in hindsight, quantifying the return period of the event is challenging~\citep{Bartusek2022, Philip2022, ZederPasche23}. 
While the impacts caused by such extreme events can be substantially large, prediction of such events is also challenging for ML models, due to the scarcity of similar events in training data. 
On the other hand, even though NWP models are more directly bound to follow physical laws, their forecast accuracy is not guaranteed either.

The heatwave impacted ecosystems, infrastructure, and human health considerably (with more than 1400 deaths), and attracted massive public attention and scientific interest \citep{neal2021PacificNorthwest2022, schumacherDriversMechanisms20212022, whiteUnprecedentedPacificNorthwest2023, rothlisbergerQuantifyingPhysicalProcesses2023}.
In the analyzed region, temperatures peaked between June 27 and June 29. We analyze the heatwave in terms of the temperature at \SI{2}{\meter} above the surface ($T_{2m}$), which is a standard variable for studying temperature extremes.

\begin{figure*}[tbp]
    \centering
    \includegraphics[width=39pc]{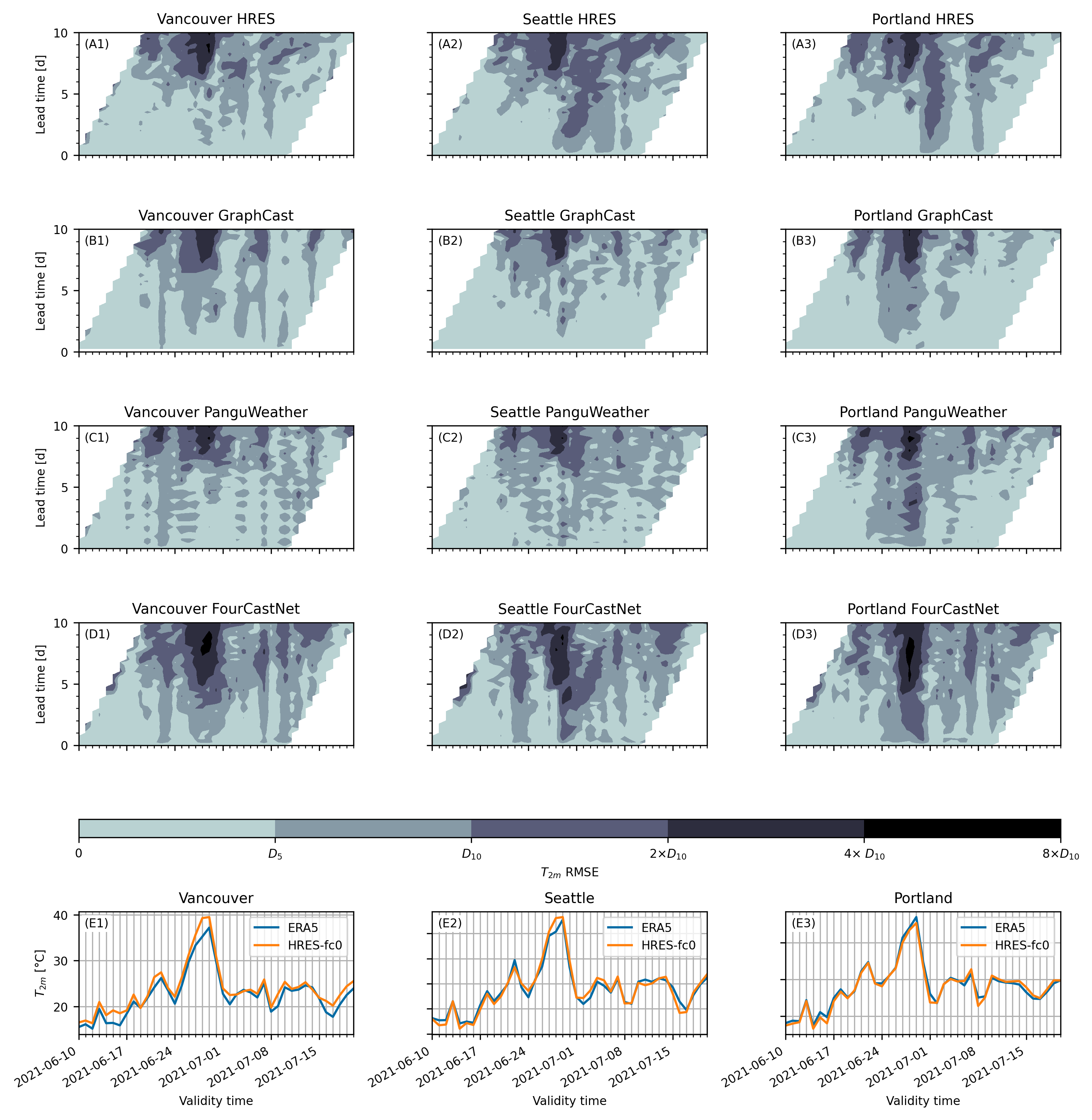}
    \caption[]{Panels (A1) to (D3): Predictability barrier plots for the grid cells closest to major cities affected by the 2021 heatwave. For HRES, HRES-fc0 is used as ground truth, for the ML models, we use ERA5 instead. In the color bar, $D_5$ and $D_{10}$ indicate long-term multi-year average HRES 5-day and 10-day prediction errors. For the computation of the $\mathrm{RMSE}$, $D_5$, and $D_{10}$ see \cref{ss:rmse}. Numerical values for $D_5$ and $D_{10}$ are given in \cref{tab:rmses}. Panels (E1) to (E3): time series of daily maximum $T_{2m}$ for the data sets used as ground truth.}
    \label{f:PNWBarrierPlots}
\end{figure*}

In the grid cells closest to three major population centers affected by the heatwave (Vancouver, Seattle, Portland), the prediction error of HRES and all tested ML models reaches at least twice the size of a typical HRES 10-day prediction error and exceeds the typical HRES 10-day error in Portland by a factor of four. 
This is consistent with the results of \citet{lin2021WesternNorth2022}, who examined the predictions of subseasonal-to-seasonal NWP models for the Pacific Northwest heatwave and found that all models failed to predict the magnitude of the heatwave for forecasts initialized on June 17, i.e., ten days before temperatures began to peak.
We visualize our prediction errors in predictability barrier plots in \cref{f:PNWBarrierPlots}, using HRES-fc0 as ground truth for the HRES forecasts and ERA5 as ground truth for the ML-based predictions. We aggregate to daily scale by computing RMSEs as described in \cref{ss:rmse}. A version of the plot showing $T_{2m, prediction} - T_{2m, ground truth}$ without this aggregation is presented in the supplementary material. Maps of the temperature anomaly patterns predicted for the peak of the heatwave for forecasts with different initialization times are shown in \cref{f:PNWAnomalies}.

FourCastNet has the largest error among all models, while the errors of PanguWeather, GraphCast, and HRES are of a similar magnitude visually (\cref{f:PNWBarrierPlots}). For all models, the prediction errors are largest during the peak of the heatwave. The predictability barrier plots exhibit prominent vertical structures (i.e., forecasts for the same validity day), suggesting that the dominant factor is the predictability of the weather situation rather than the forecast initialization. However, HRES seems to also exhibit hints of diagonal error structures. This structural difference in error patterns is further discussed in \cref{ss:na_winterstorm}, where it is more significant.

For HRES, the prediction errors in the first days of July 2021, when temperatures started to fall again, are larger than for PanguWeather and GraphCast, especially for the grid cells closest to Seattle and Portland. For long lead times, however, the HRES errors reach their largest values during the heatwave peak around June 27-29 in all three grid cells. The predictability barrier plots for PanguWeather appear very patchy, likely due to the ``hierarchical temporal aggregation strategy'' of PanguWeather (described in \cref{ss:methods}).

The best and worst performing models across various lead and validity times are visualized in \cref{f:PNWBestWorst}. Conclusions match those from \cref{f:PNWBarrierPlots}; FourCastNet has the largest errors during the heatwave, and HRES has comparatively high errors after the peak of the heatwave. During many of the time steps, especially for short lead times, GraphCast and HRES yield the smallest error. However, there is no clear best-performing model overall.

\begin{figure*}[tb]
    \centering
    \includegraphics[width=0.8\textwidth]{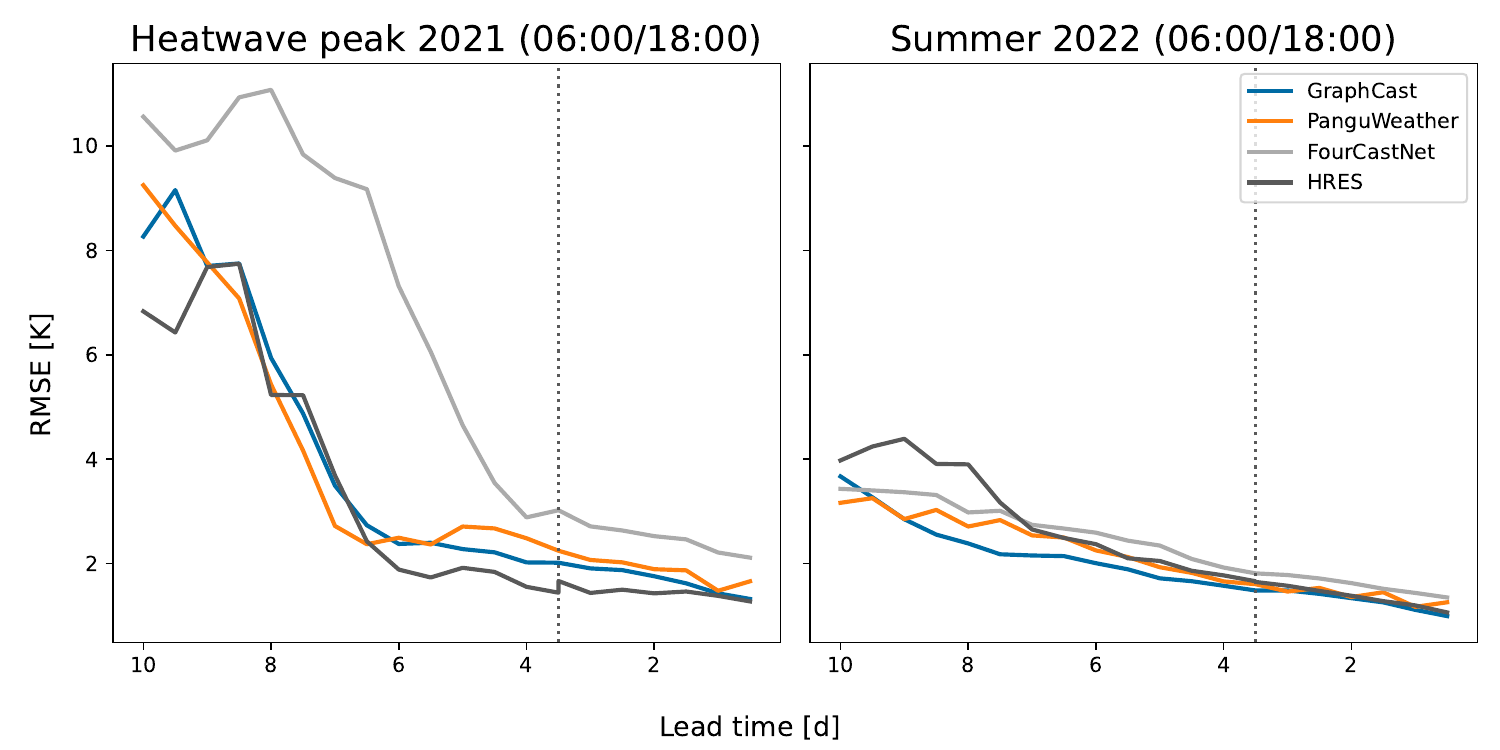}
   \caption[]{Evolution of the $T_{2m}$ prediction RMSE with lead-time for the three ML models and HRES in the event region during the peak of the heatwave (June 27--29 2021, left) compared to summer 2022 as a baseline (June~20--July~10, right). Observations in the considered box region, $45^\circ$--$52^\circ$N, $119^\circ$--$123^\circ$W, are weighted to correct for differences in grid-cell area. ML models use 06:00/18:00 UTC initial conditions and evaluation times only, and the HRES forecasts use the mixed initialization described in \cref{ss:init-time} after \qty{3.75}{\day} (dotted line).}
    \label{f:PNWrmseBaseline}
\end{figure*}

To assess the models' performance in predicting the extreme event, we compute the forecast RMSEs of all models during the peak of the heatwave and compare them to the RMSEs during the summer of 2022, a baseline year without extreme heatwaves in the region. We vary the lead time and study the event in the region defined by~\citet{Philip2022}. The RMSE aggregation here follows \citet{lamLearningSkillfulMediumrange2023}: it includes latitude-based weights, and only forecasts initialized at 06:00/18:00 UTC and lead times in multiples of \qty{12}{\hour} are considered to ensure equal assimilation windows between ERA5 and HRES-fc0. 
The results, shown in \cref{f:PNWrmseBaseline}, again highlight the difficulty of predicting the extreme temperatures during the event: for lead times beyond one week, all models perform substantially worse than for the summer 2022 baseline. More precisely, all ML models perform up to at least three times worse and HRES up to two times worse. 
Given the small sample size, these numerical values should be interpreted with care, however.
We also observe that for lead times up to \qty{6.5}{\day}, the forecast errors of HRES are smaller than ML models, contrary to their performance in the baseline year.
This might be a consequence of extrapolation, as discussed in the introduction. 
As the evaluated baseline period is rather short, mainly for computational reasons, the baseline might not precisely represent the typical performance. 
However, the baseline results are in line with other studies \citep{lamLearningSkillfulMediumrange2023} and additional baseline data from the year 2020 (\cref{f:PNW_RMSE2020_2022} in \cref{As:PNW}).

A further analysis focusing on the spatial aspect of the event is presented in Section~\ref{sups:PNW} of the Supplementary Materials. A main conclusion is that FourCastNet under-predicts the area in which temperature anomalies exceed a given threshold, while in some PanguWeather forecasts, the area predicted to exceed the thresholds is too large.

\subsection{2023 South Asian Humid Heatwave} %
\label{ss:sa_humid_heatwave}

In April 2023, high temperature and humidity levels were reached simultaneously in South Asia~(\cref{f:IntroPlots}B). The human tolerance to high temperatures decreases with increasing humidity, mainly due to the inability of the body to self-regulate its temperature through transpiration. Heat stress associated with this type of event can therefore be particularly harmful to human health \citep{buzanMoistHeatStress2020, loOptimalHeatStress2023}.

The heat index ($HI$) is an impact metric quantifying this hazard to human health. It estimates the apparent temperature (i.e., how hot the temperature feels) for given values of temperature ($T_{2m}$) and relative humidity ($RH$). While many metrics have been proposed to combine the influence of these two variables \citep{loOptimalHeatStress2023}, we follow \cite{zachariahExtremeHumidHeat2023a}, who employ the modified version of the heat index \citep{rothfuszHeatIndexEquation1990} used by the NOAA Weather Prediction Center in an attribution study on the 2023 South Asian humid heatwave. The detailed computations, including information on how we convert predicted specific humidity to relative humidity, are given in \cref{ss:heat-index}.

Following \cite{zachariahExtremeHumidHeat2023a}, we focus on two study regions in South Asia: Laos-Thailand (for which results are presented in Section~\ref{sups:AHH} of the Supplementary Materials), and India-Bangladesh. For the latter, a sub-region with dry and semi-arid climate is excluded from the analysis (see \cref{ss:shape_files} for details). We select a temporal range of April~17--20, 2023 (UTC time, inclusive range) for the India-Bangladesh region, corresponding to the period in which the heat stress peaked.

With existing ML weather prediction models, ${HI}$ at the surface cannot be calculated correctly because humidity is only modeled at higher pressure levels, and none of the models predict a variable that could enable the calculation of relative humidity at the surface level. This presents a strong limitation on the utility of ML models in forecasting humid heat waves. While GraphCast and PanguWeather forecast variables at the \qty{1000}{\hecto\pascal} level, FourCastNet predictions only include variables at pressure levels starting from \qty{850}{\hecto\pascal}. In the following, we exclude FourCastNet from the analysis and use relative humidity at the \qty{1000}{\hecto\pascal} level as an approximation for humidity at the surface.

\begin{figure*}[tbp]
    \centering
    \includegraphics[width=27pc]{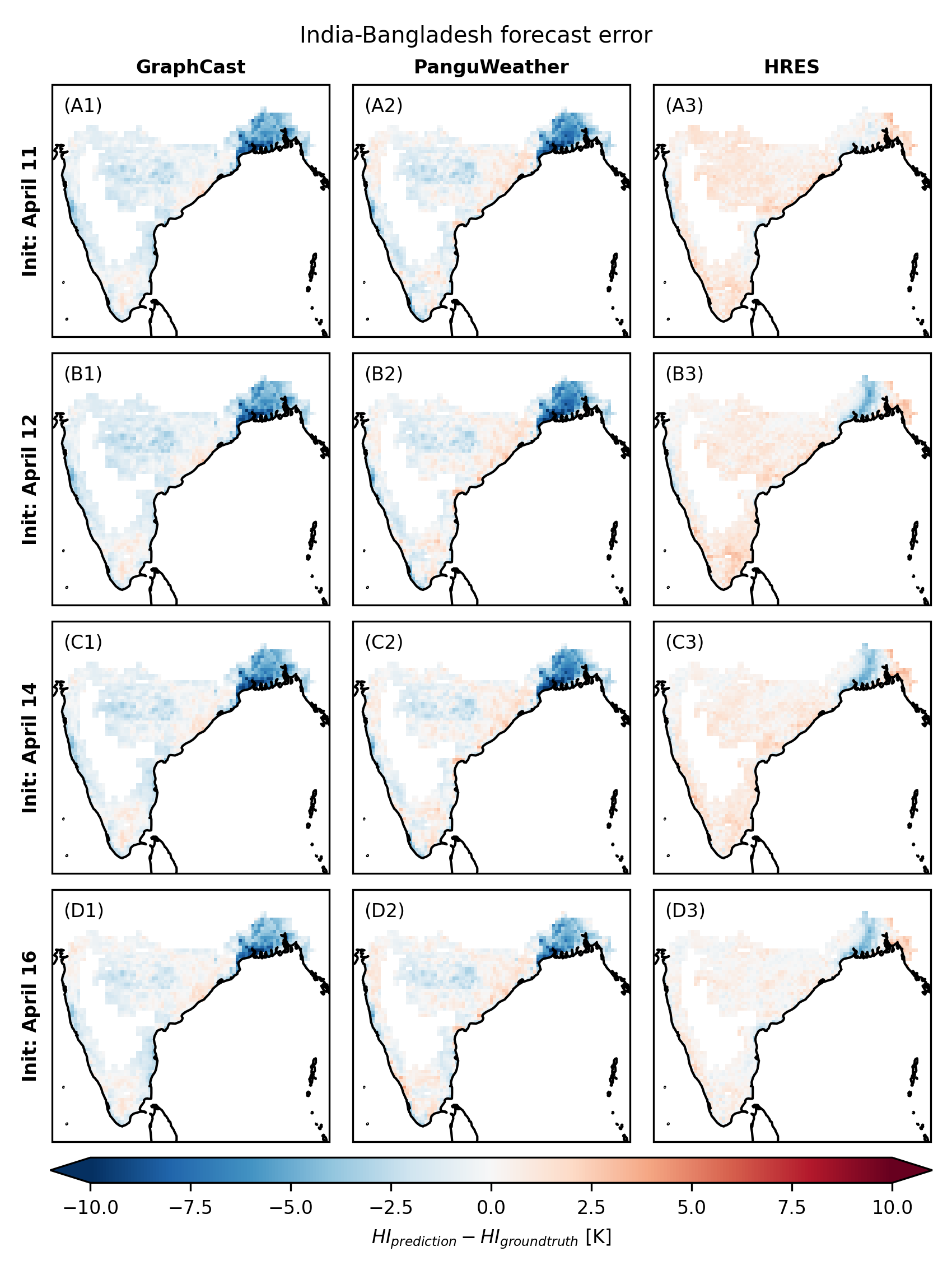}
    \caption[]{Error of the ${HI}$ prediction, for the time step of each day during which $HI$ peaked in the ground truth data set, averaged over April~17--20, 2023. For all forecasting methods and ground truth data sets, ${HI}$ is computed using ${RH}_{\qty{1000}{\hecto\pascal}}$ rather than the value at the surface.}
    \label{f:AHHspatialInit}
\end{figure*}

The ${HI}$ prediction error during the peak of the heatwave in the India-Bangladesh region is shown in \cref{f:AHHspatialInit}. For each day, we select the time when the ground truth $HI$ is maximal, and then average the errors over April~18--20. 
In all cases, the predicted ${HI}$ is computed using ${T}_{2m}$ and ${RH}_{\qty{1000}{\hecto\pascal}}$. This setup is the simplest possible substitute that forecasters could use with the ML models available at the time of writing.
The forecasts show deviations from the ground truth data sets, especially over Bangladesh. The under-prediction for ML models is stronger than for HRES. Looking at the prediction errors of ${RH}_{\qty{1000}{\hecto\pascal}}$, we find a matching pattern: predictions for relative humidity over Bangladesh are too low, especially for PanguWeather (\cref{f:AHHspatialHumidityPressureLevel}). For $T_{2m}$, the values at the time of the $HI$ peak are mostly smaller than the corresponding ground truth for the ML methods, while HRES $T_{2m}$ predictions are larger than the HRES-fc0 ground truth (\cref{f:AHHspatialHumidityPressureLevel}). 

For HRES, HRES-fc0, and ERA5 it is possible to compute relative humidity at the surface level ($RH_{sfc}$) from 2m temperature and 2m dewpoint temperature (see \cref{ss:heat-index}). We found rather large differences between the ground truth data sets ERA5 and HRES-fc0 for the studied event however, therefore we used ${RH}_{\qty{1000}{\hecto\pascal}}$ in the computations for \cref{f:AHHspatialInit}.

\begin{figure*}[tb]
    \centering
    \includegraphics[width=27pc]{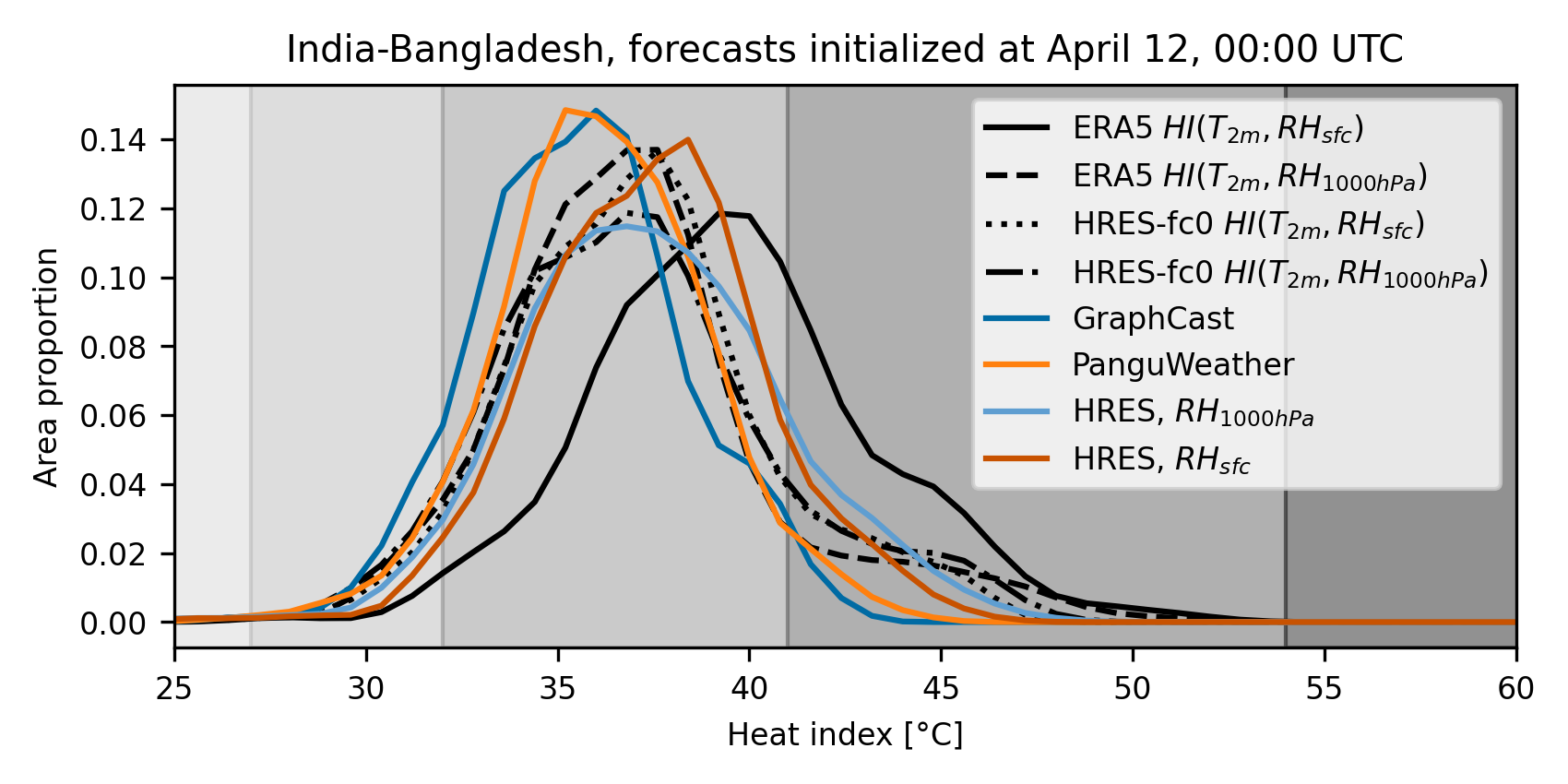}
    \caption[]{Proportion of area in study region with given mean daily maximum heat index during April~17–20, 2023, computed using area-weighted kernel density estimation. Shaded areas in the background indicate threat levels (see \cref{ss:heat-index}). Light gray to dark gray: low risk, caution, extreme caution, danger, extreme danger. Compared are distributions resulting from forecasts initialized 6  days prior to the start of the event and different ground truths: ERA5 and HRES-fc0, each in two versions of computing the heat index either using $RH_{sfc}$ or using the substitute $RH_{\qty{1000}{\hecto\pascal}}$. For HRES forecasts, we show versions computed with $RH_{\qty{1000}{\hecto\pascal}}$ and $RH_{sfc}$ as well.}
    \label{f:AHHHIAreaIndiaBangladesh}
\end{figure*}

Large fractions of the India-Bangladesh region experienced a mean daily maximum $HI$ during April~17--20 2021 that falls in the ``extreme caution'' or ``danger'' category (\cref{f:AHHHIAreaIndiaBangladesh}, see \cref{tab:hi_categories} for the definition of the categories). In \cref{f:AHHHIAreaIndiaBangladesh}, the $HI$ distribution computed from ERA5 data using the input variables $T_{2m}$ and $RH_{sfc}$ differs strongly from the other ground truth data sets, mainly due to ERA5 $RH_{sfc}$ values being higher during daily maximum $HI$. This may be a consequence of differences in the assimilation procedure used to produce the ground truth data.
The ML-based ${HI}$ forecasts (computed using ${RH}_{\qty{1000}{\hecto\pascal}}$) underestimate the ERA5 $HI$ values both when $RH_{\qty{1000}{\hecto\pascal}}$ or $RH_{sfc}$ is used. This is especially the case for high values of $HI$. Results for the Laos-Thailand region are also in line with these findings (see Section~\ref{sups:AHH} in supplementary information). 

\Cref{f:AHHcategoriesForecast} visualizes the heat index forecast for the peak of the heatwave by forecasts with different initialization times (in terms of threat categories, see \cref{tab:hi_categories}). Results match the other analyses in this section.

\subsection{2021 North American Winter Storm}
\label{ss:na_winterstorm}

While heatwaves often receive a lot of media attention, especially in light of anthropogenic climate change, cold spells are also hazardous. Under the current climate, cold extremes lead to more human deaths overall than hot extremes \citep{gasparriniMortalityRiskAttributable2015}.
In mid-February 2021, a winter storm hit large parts of the United States, Northern Mexico, and Canada (\cref{f:IntroPlots}C). Rapidly falling temperatures were accompanied by snow, sleet, freezing rain, and strong winds, causing damages to human livelihood and infrastructure \citep{national2021valentine}. In Texas, which was strongly affected by the event, pipes burst, interrupting the water distribution, and energy infrastructure failed, resulting in power outages and ordered rolling blackouts. Impacts were amplified by inadequate wintering of energy infrastructure \citep{gruberProfitabilityInvestmentRisk2022}.

While it would be interesting to look at a metric that directly relates to vulnerabilities of the Texas power grid, defining such a metric is not straightforward. Therefore, we restrict our analyses to $T_{2m}$ and the wind chill index $T_{wc}$ as defined in \citet{osczevskiNEWWINDCHILL2005}. The wind chill index is a metric that describes the apparent temperature in the presence of wind. It is defined as 
\begin{equation}
    \label{eq:wind_chill}
    T_{wc} = 13.12 + 0.6215 T_{2m} - 11.37 v^{0.16} + 0.3965 T_{2m} v^{0.16},
\end{equation}
where $T_{wc}$ and $T_{2m}$ are given in \unit{\celsius}, and the wind-speed at \qty{10}{\meter} height $v$ in kilometers per hour (computed from the horizontal wind components). The formula was obtained by modeling heat transfer from the human body to the atmosphere \citep{osczevskiNEWWINDCHILL2005}. Note that $T_{wc}$ is only defined for temperatures below \qty{10}{\celsius} and wind speeds above \qty{4.8}{\km\per\hour}. 
In a particularly affected Texas city, College Station, HRES-fc0 and ERA5 closely follow observational records for both $T_{wc}$ and $T_{2m}$ (see \cref{f:AWSCollegeStation} in \cref{As:AWS}), demonstrating the suitability of those datasets for this index. %

\begin{figure*}[tbp]
    \centering
    \includegraphics[width=27pc]{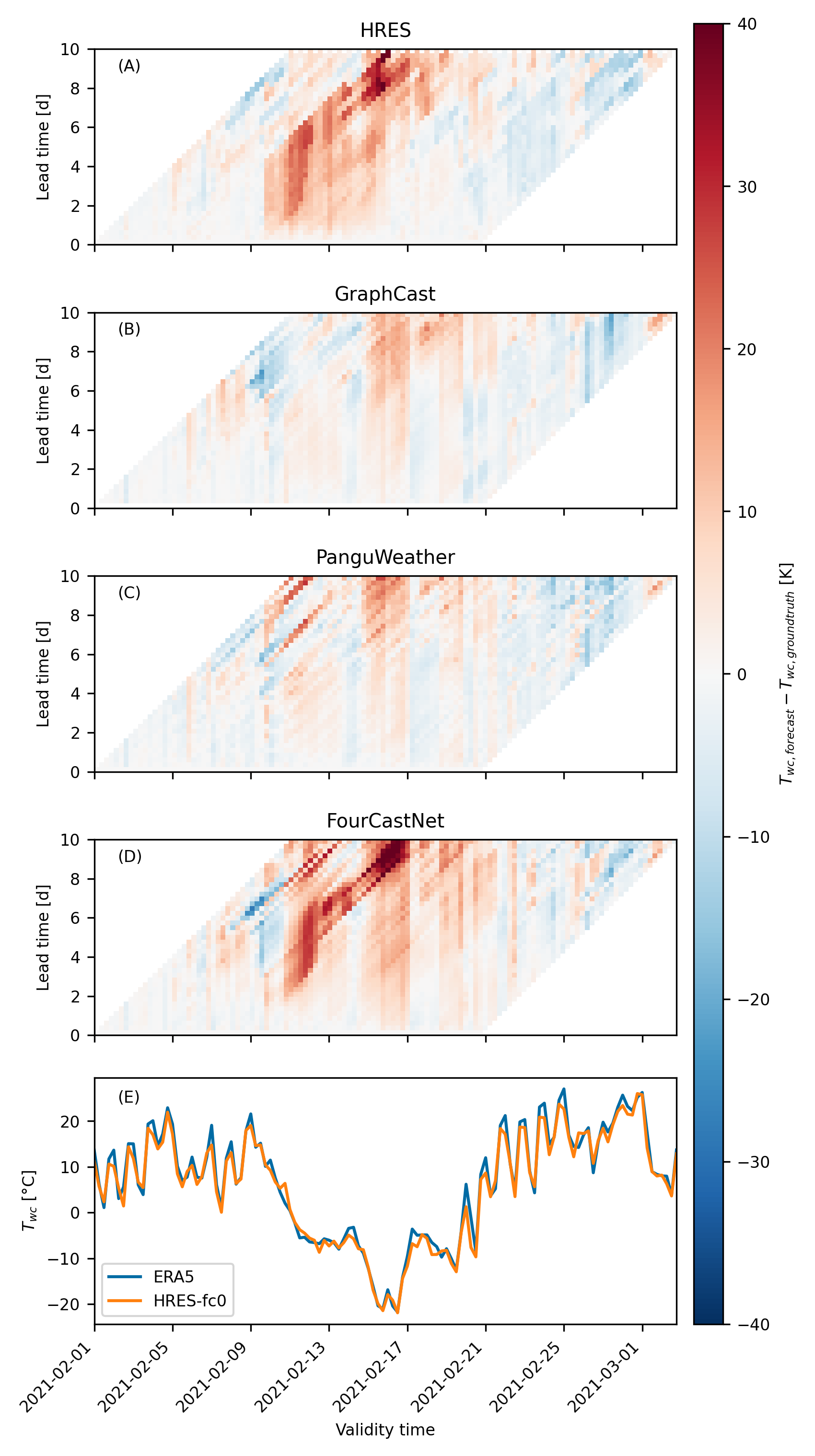}
    \caption[]{Panels (A) to (D): ${T}_{wc}$ prediction errors for different validity and lead times. Data from grid cell closest to College Station, Texas. Times and dates are given in UTC. Panel (E): time series of the ground truth data sets used in the computation: ERA5 is used for the ML forecasts, HRES-fc0 is used for HRES.}
    \label{f:AWSPredBar}
\end{figure*}

Looking at predictions of ${T}_{wc}$ for the grid cell closest to College Station in \cref{f:AWSPredBar}, one can see that all models struggle to predict the minimum wind chill index, with forecast errors being largest for FourCastNet (sometimes exceeding \qty{40}{K} at minimum $T_{wc}$ for large lead times). In general, errors are larger for the winter storm than for the Pacific Northwest heatwave, which might be due to a potential seasonality in prediction errors \citep[as suggested by Figure 2 in][]{benBouallegueRiseDataDrivenWeather2024} or, simply, to the different nature of the events.
Errors for PanguWeather and GraphCast are substantially lower than for HRES, especially between February 9 and February 17, after the peak of the winter storm. 

The vertical structures in the plot (which are particularly prominent for GraphCast and PanguWeather on February~15--17) hint at the difficulty of predicting weather situations, while the diagonal structures (strong for FourCastNet and HRES) suggest variation between individual forecasts caused by their initial conditions. In general, HRES seems to produce stronger diagonal error structures, while the ML models tend to exhibit vertical error patterns. While the data filling we use to extend HRES forecasts might affect this finding, it could also be a result of fundamental differences between ML-based and NWP forecasts. Assuming a very extreme event that is easily predictable following physical laws, the predictions of HRES would steadily improve when approaching the event. ML methods, however, might not be able to extrapolate to such extreme conditions, even at very short lead times, resulting in vertical error patterns. 
On the other hand, if an extreme event that is difficult to predict with (first-order) physical laws is somewhat hidden in the atmospheric state of the initial-conditions day, HRES would have trouble forecasting both the event and its buildup, leading to a diagonal pattern. Then, when the event becomes more apparent from the conditions, the prediction will improve. The ML methods might be able to model such `hidden' (second-order) conditions better because of their flexibility, leading to less strong diagonal patterns. %

\begin{figure*}[tbp]
    \centering
    \includegraphics[width=33pc]{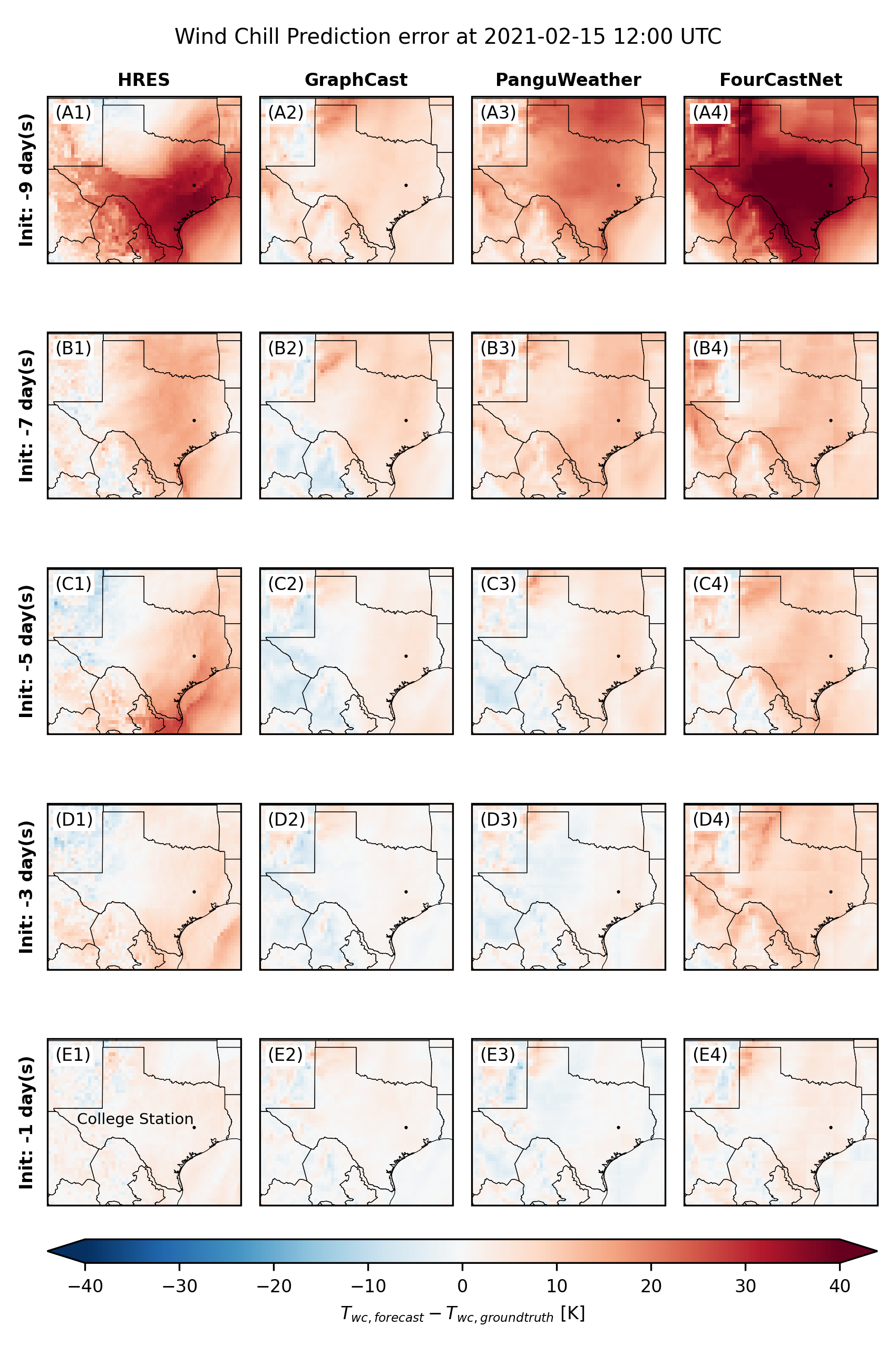}
    \caption[]{Errors of ${T}_{wc}$ forecasts for 12:00 UTC, February 15 2021 (6:00 am Houston time). The ground truth used to compute results for the ML forecasts is ERA5, while HRES-fc0 is used for HRES.}
    \label{f:AWSPredictionErrorPatterns}
\end{figure*}

When the predicted temperatures are too high, or the predicted wind speeds too low, the thresholds in the definition of $T_{wc}$ are not exceeded, and $T_{wc}$ is thus not defined. 
This is the case during the wind chill minimum between February~15 and February~17, even though these were the most hazardous days. In \cref{f:AWSPredBar,f:AWSPredictionErrorPatterns} we ignore the thresholds in the definition and still compute the $T_{wc}$ expression for visual clarity.

Patterns in the prediction of ${T}_{2m}$ look similar to those of ${T}_{wc}$ (\cref{f:AWSPredBarrT2m}), with FourCastNet errors being largest, and HRES errors during the event larger than PanguWeather and GraphCast errors. For the surface windspeed, which also enters Equation \eqref{eq:wind_chill}, the patterns are not as clear (\cref{f:AWSPredBarrWindspeed}). Therefore, the ${T}_{wc}$ errors seem to be dominated by ${T}_{2m}$.

\Cref{f:AWSPredictionErrorPatterns} depicts the forecast errors in a bounding box around Texas (\qtyrange{107}{93}{\degree W}, \qtyrange{25}{37}{\degree N}) for 12:00 UTC on February 15, 2021 (6:00 am, Houston time), when the average wind chill index in the box reached a minimum in the ground truth data.
For long lead times, forecast errors of GraphCast and PanguWeather seem smaller than those of FourCastNet and HRES, although all forecasts appear to be too warm.
One can also notice a slightly ``patchy'' structure in the FourCastNet predictions, with notable discontinuities between 8x8 patches. 8x8 is the patch size used internally by FourCastNet.

\section{Discussion and Conclusions}
\label{s:discussion-conclusion}

The three case studies conducted highlight different aspects of comparison between the ML models GraphCast, PanguWeather, FourCastNet, and the NWP system HRES.
For the 2021 Pacific Northwest heatwave, predictions of PanguWeather and GraphCast maintained comparable quality to HRES in terms of the evaluated metrics. 
However, for short lead times, HRES showed smaller forecast errors in both the predictability barrier plot and RMSE plot than ML models, contrary to their performances in the baseline summer of 2020 or 2022, indicating that ML models might be more severely impacted by the extrapolation to those extreme conditions. 
We also observe that HRES has more difficulties predicting the sharp drop in the temperature after the peak of the heatwave than ML models.
When analyzing the South Asia humid heatwave substituting $RH_{sfc}$ with $RH_{\qty{1000}{hPa}}$, prediction errors show spatial patterns with the highest danger levels over Bangladesh being underestimated by the ML models.
For many lead times and initial conditions, the North American Winter Storm is forecast more accurately by PanguWeather and GraphCast than by HRES. 
From those predictions, we observe structurally different error patterns: HRES and FourCastNet are potentially more impacted by subtle signals in the initial conditions than GraphCast and PanguWeather, leading to errors that build up before the event. 
We emphasize that our findings are limited to the three case studies, and more systematic analyses need to be conducted to reach definitive conclusions about general extreme weather event forecasts.

None of the ML models predict a variable that enables the computation of surface-level humidity. This would have allowed us to better study the effects of the 2023 humid heatwave in South Asia, as surface humidity alters the effect of temperature on the human body. Looking at substitute variables, ML models seem to perform worse for this event overall, potentially due to extrapolation. Whether this effect persists for ML models that do predict surface humidity remains to be answered in future research. The rather large differences in relative humidity at the surface level between the ERA5 data set and HRES-fc0 complicate the estimation of ``true'' heat index forecasting errors. One way to resolve this would be to directly compare against station observations.

Comparing forecast systems on only a subset of extreme events incurs the danger of favoring alarmist forecasts, a phenomenon termed the ``forecaster's dilemma'' \citep{Lerch2017}. This is a general problem of case studies and even applies to a broader class of analyses. %
Such results should not be used alone to judge the overall quality of weather forecasting systems. The existing literature on evaluating ML-based weather forecasts described in \cref{s:introduction} can be combined with our findings to obtain a more complete picture.

Our study only uses single forecasts and disregards probabilistic forecasting. In NWP, forecast uncertainty is accounted for by running ensemble forecasts. While including NWP ensemble forecasts is possible, this would have caused further complications in the analyses, e.g. due to differing model resolutions. Furthermore, producing ensemble forecasts with the given ML models is non-trivial. Attempts have been made, e.g., by perturbing initial conditions or model parameters \citep{weynSubSeasonalForecasting2021, biAccurateMediumrangeGlobal2023, bülte2024uncertainty}, but problems capturing the right scaling of uncertainties have been found in the literature. \citet{selzCanArtificialIntelligence2023} investigated PanguWeather and found that the error growth for small perturbations is too small (``no butterfly effect''). Recently, \citet{priceGenCastDiffusionbasedEnsemble2023} explored generative modeling to obtain better ensembles. Because of the generative training objective, these models can better capture the spectrum of the weather at long lead times, avoiding the over-smoothing that occurs for autoregressive models like GraphCast, PanguWeather, and FourCastNet. For the evaluation of generative ML-based weather forecast models, proper scoring rules \citep{gneitingProbabilisticForecasting2014}, like the class described by \citet{Allen2023} for probabilistic forecasts, will be an important analysis tool.
The comparison of ML models with HRES is also limited by differences between the ground truth data sets ERA5 and HRES-fc0 (differing assimilation times, short forecasts for 06:00/18:00 UTC initializations). However, this does not affect the comparison between the three ML models. ML weather prediction models are typically trained using ERA5 data, which does not correspond to an ``operational setting'' and complicates the comparison with HRES. While ML models could be trained or fine-tuned with HRES-fc0 data directly, the IFS version used to produce the forecasts varies over time, therefore characteristics and biases of the ``ground truth'' HRES-fc0 would also vary.

Most ML models employ a large autoregressive time step (6h for GraphCast and FourCastNet). This coarse temporal resolution might affect the forecast of impacts for which the daily maximum or minimum is relevant, such as short-term heat stress peaks or severe wind gusts. The most extreme values might be missed due to an unfortunate combination of forecast time step and daily cycle or event time. Some important variables for impact assessments are forecast by few or no ML models. These include humidity at the surface, solar radiation reaching the surface (potentially relevant for solar energy production forecasts), and precipitation. While some ML models (including GraphCast and FourCastNet) do predict precipitation, authors have advised caution in the interpretation of this variable, citing issues with the ERA5 precipitation ground truth \citep{laversEvaluationERA5Precipitation2022}.

While case studies can only provide anecdotal evidence, testing ML models under individual extreme events can reveal unexpected deficiencies (or advantages) of these models in comparison to well-established techniques. The rather small number of meteorological variables predicted by ML models, as well as the available forecast lead time, limits the types of impactful extreme events that can be studied for these models. While longer forecasts would be interesting and would allow the study of more complex types of extreme events \citep{zscheischlerTypologyCompoundWeather2020}, one would likely need to include new processes and variables into the models, such as feedback from soil moisture and the influence of sea-surface temperatures. 

Non-linear combinations of predicted output variables (e.g., wind chill, see \cref{eq:wind_chill}) have the potential to reveal weaknesses of ML models; \citet{priceGenCastDiffusionbasedEnsemble2023} investigated horizontal surface wind speed (a non-linear function of the horizontal wind components) and found that GraphCast tends to perform worse in terms of this combined metric than for the individual components. They hypothesized that this might be due to the tendency of a certain type of ML architecture to predict close to the mean under forecast uncertainty and the non-commutativity of non-linear function applications and averaging.
However, in our case studies ($T_{wc}$ during the 2021 North American winter storm and $HI$ during the 2023 South Asia humid heatwave), the large differences in prediction errors for individual input variables ($T_{2m}$ during winter storm, relative humidity during humid heatwave) and the necessity of having to substitute relative humidity at the surface level seem to outweigh this effect. Nevertheless, the described problem is an interesting target for future work, as impacts often are not simply determined by linear combinations of the variables predicted by the ML models. \citet{priceGenCastDiffusionbasedEnsemble2023} suggest using generative modeling to overcome this systematic problem.

For theoretically-justified extrapolation to extremes and when interested in risk assessment, a natural approach is the use of extreme value statistics~\citep{Coles2001}. 
Recently, various approaches have combined machine learning and extreme value statistics to improve predictive extrapolation of extreme risk for the predicted variable \citep{Pasche2024,RichardsNNfire,gbex,CisnerosNN,AlloucheNN,erf}. 
Methods for extrapolation in the predictor space also exist but require stronger dependence assumptions \citep{ChenEngression,PfisterExtrapol}.
Including physical domain knowledge in ML-based models, for example through architectural restrictions of explicit equations, could be another approach to improving generalization~\citep{kochkovNeuralGeneralCirculation2024}. 

Releasing raw predictions instead of aggregates or summaries, or even pre-trained models, is valuable \citep{burnellRethinkReportingEvaluation2023}. As considering all metrics that stakeholders deem important during model development and testing is challenging or impossible, releasing the full predicted data or trained models allows assessing domain-specific model skill even after the development of the model. WeatherBench 2 \citep{raspWeatherBenchBenchmarkNext2024} already partially addressed this point. Building a continuously-updated database of extreme event case study set-ups \citep[similar to ECMWF's Severe Event Catalogue, ][]{SevereEventCatalogue}, including domains and impact metrics, might be a valuable contribution to existing literature. A focus on re-usability for new models would be important, potentially through the integration of a framework like WeatherBench~2. One could hypothesize that the selection of extreme events based on their real-world impacts leads to a ``selection bias'', as the larger impacts might have been caused by poor forecasts by operational models, potentially resulting in a biased estimate of the relative performance of HRES.

The evaluation of ML models typically focuses on meteorological variables. Putting a stronger focus directly on impacts has the potential to improve the practical value of ML models. To find suitable impact metrics, researchers could, for instance, look at warnings issued by weather services and analyse how warnings based on NWP compare to forecasts using ML models. Coupling ML weather forecasts to impact models, such as models for floods \citep{nearingGlobalPredictionExtreme2024}, crop loss, or fires, might also be valuable, although the analysis would then also depend on the impact model's fidelity. While ML models have shown impressive skill in forecasting key meteorological variables, it is worthwhile investigating whether their predictions can lead to similarly impressive results when assessing impacts.

\newpage

\appendix

\vspace{10mm}

\setcounter{section}{0}
\setcounter{subsection}{0}
\setcounter{equation}{0}
\setcounter{figure}{0}
\setcounter{table}{0}

\renewcommand{\thesection}{A.\arabic{section}}
\renewcommand{\thesubsection}{A.\arabic{section}.\arabic{subsection}}
\renewcommand{\theequation}{A.\arabic{equation}}
\renewcommand{\thefigure}{A.\arabic{figure}}
\renewcommand{\thetable}{A.\arabic{table}}

\renewcommand{\theHsection}{A.\thesection}
\renewcommand{\theHsubsection}{A.\thesubsection}
\renewcommand{\theHequation}{A.\theequation}
\renewcommand{\theHfigure}{A.\thefigure}
\renewcommand{\theHtable}{A.\thetable}

\section{Further Details and Analysis of the 2021 Pacific Northwest Heatwave}
\label{As:PNW}
\subsection{Computation of the Root Mean Squared Error}
\label{ss:rmse}
When computing the root mean squared error ($\mathrm{RMSE}$), we follow \citet{raspWeatherBenchBenchmarkNext2024}, who choose to include the average over initialization time steps inside the square root of the $\mathrm{RMSE}$ formula, in opposition to earlier works in the field \citep{lamLearningSkillfulMediumrange2023, raspWeatherBenchBenchmarkData2020}. For a given variable $x$ (say 2m temperature $T_{2m}$), its prediction $\hat{x}$, and a fixed prediction lead time $\tau$, we compute its $\mathrm{RMSE}(\tau)$ as
\begin{equation}
    \label{eq:RMSE}
    \mathrm{RMSE}_{\mathcal{T}, \mathcal{P}}(\tau) =  \sqrt{\frac{1}{|\mathcal{T}|}\frac{1}{|\mathcal{P}|} \sum\limits_{t_0 \in \mathcal{T}}\sum\limits_{i \in \mathcal{P}} a_i \left(\hat{x}_i^{t_0+\tau} - x_i^{t_0+\tau}\right)^2},
\end{equation}
where $\mathcal{T}$ and $\mathcal{P}$ are the sets of initialization times and grid points of interest, $|\mathcal{T}|$ and $|\mathcal{P}|$ are their cardinalities, $\hat{x}_i^{t_0+\tau}$ is the forecast of the variable $x$ at grid cell $i$ and time $t_0+\tau$, with forecast initialized at time $t_0$, and $x_i^{t_0+\tau}$ is the ground truth for the same grid cell and time step $t_0+\tau$.
The weight $a_i = \cos(\text{lat}_i *\pi/180)$ is proportional to the area of the latitude-longitude grid cell $i$ and varies with the latitude of $i$, $\text{lat}_i$, and $\{a_i, i \in \mathcal{P}\}$ are normalized to have unit mean across the included grid cells. To compute the average RMSE for a given day and lead time $\tau$, we include only initialization times $\tilde{\mathcal{T}} \subseteq \mathcal{T}$ such that $t_0 + \tau$ falls on the day of interest for $t_0 \in \tilde{\mathcal{T}}$.

In \cref{f:PNWBarrierPlots}, two contours are determined by the long-term average HRES performance for \qty{120}{\hour} ($D_5$) and \qty{240}{\hour} ($D_{10}$) forecasts. To estimate these values, we use all HRES forecasts provided by WeatherBench~2 \citep{raspWeatherBenchBenchmarkNext2024}, which at the time of the writing contain only initializations at 00/12 UTC between January 1, 2016, and January 10, 2023. We use HRES-fc0 as the ground truth and only consider predictions for days within a window size of 45 days around the day-of-year of June 28, 2021. Numerical values for the grid boxes closest to the three investigated cities are provided in \cref{tab:rmses}.

\begin{table*}[tb]
    \centering
    \begin{tabular}{c||c|c|c}
        & Vancouver & Seattle & Portland \\
        \hline \hline
        $D_5$ ($T_{2m} \mathrm{RMSE}$ at day 5) &  \qty{1.53}{\kelvin} & \qty{1.61}{\kelvin} & \qty{1.93}{\kelvin} \\
        $D_{10}$ ($T_{2m} \mathrm{RMSE}$ at day 10) &  \qty{2.80}{\kelvin} & \qty{2.82}{\kelvin} & \qty{3.73}{\kelvin} \\
    \end{tabular}
    \caption{Long-term average RMSE values of HRES predictions for lead times of 120 hours ($D_5$) and 240 hours ($D_{10}$), computed as described in \cref{ss:rmse}.}
    \label{tab:rmses}
\end{table*}

\subsection{Additional Figures}
\begin{figure}[tb] 
    \centering
    \includegraphics[width=0.4\textwidth]{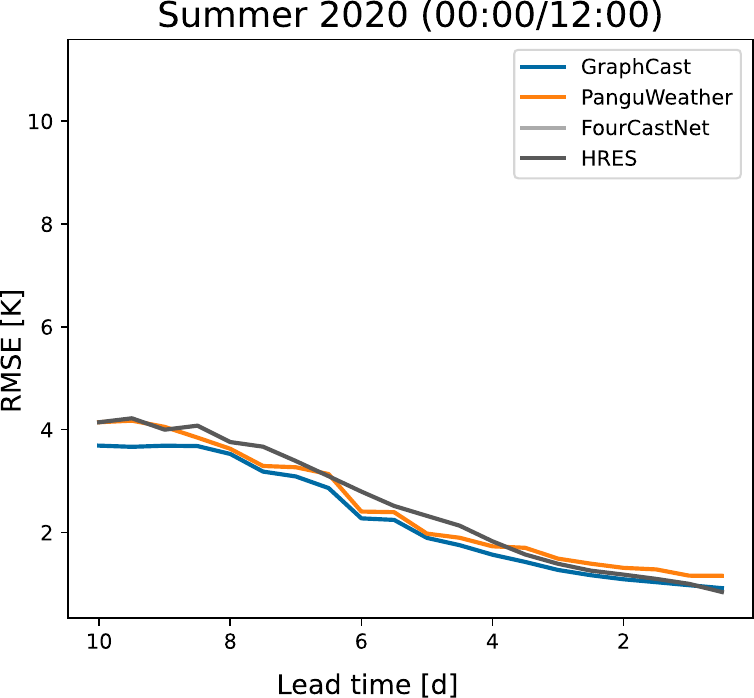}
    \caption[]{Evolution of the $T_{2m}$ prediction RMSE with lead-time for the three ML models and HRES in the 2021 Pacific Northwest heatwave region during the summer 2020 (June~1--July~31) as a baseline year. Observations in the considered box region, $45^\circ$--$52^\circ$N, $119^\circ$--$123^\circ$W, are weighted to correct for differences in grid-cell area. Both ML models and HRES use 00:00/12:00 UTC initial conditions and evaluation times only, as opposed to \cref{f:PNWrmseBaseline}, since the forecast were downloaded from WeatherBench 2 for computational reasons, where only these initializations are available.}
    \label{f:PNW_RMSE2020_2022}
\end{figure}

\begin{figure*}[tbp] 
    \centering
    \includegraphics[width=0.8\textwidth]{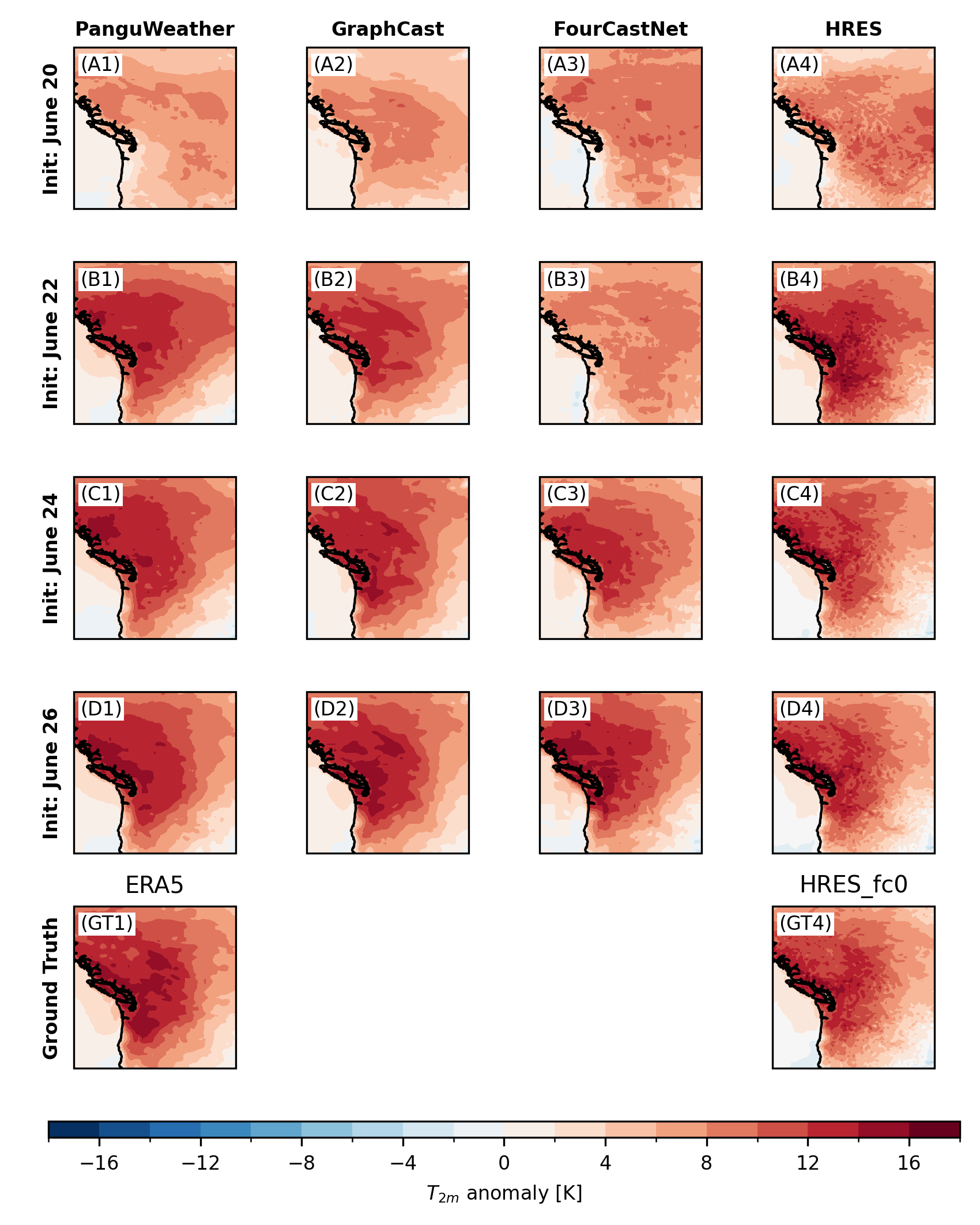}
    \caption[]{Average temperature anomaly predicted for June 27--June 29 2021 (inclusive). All anomalies including those for HRES are calculated with respect to the ERA5 climatology given in \citet{raspWeatherBenchBenchmarkNext2024}. The fact that HRES anomalies are computed against ERA5 data explains the patchy small-scale structure visible in the HRES panels. Forecasts are initialized at 00 UTC on the day specified in the row title.}
    \label{f:PNWAnomalies}
\end{figure*}

\begin{figure*}[tbp]
    \centering
    \includegraphics[width=0.8\textwidth]{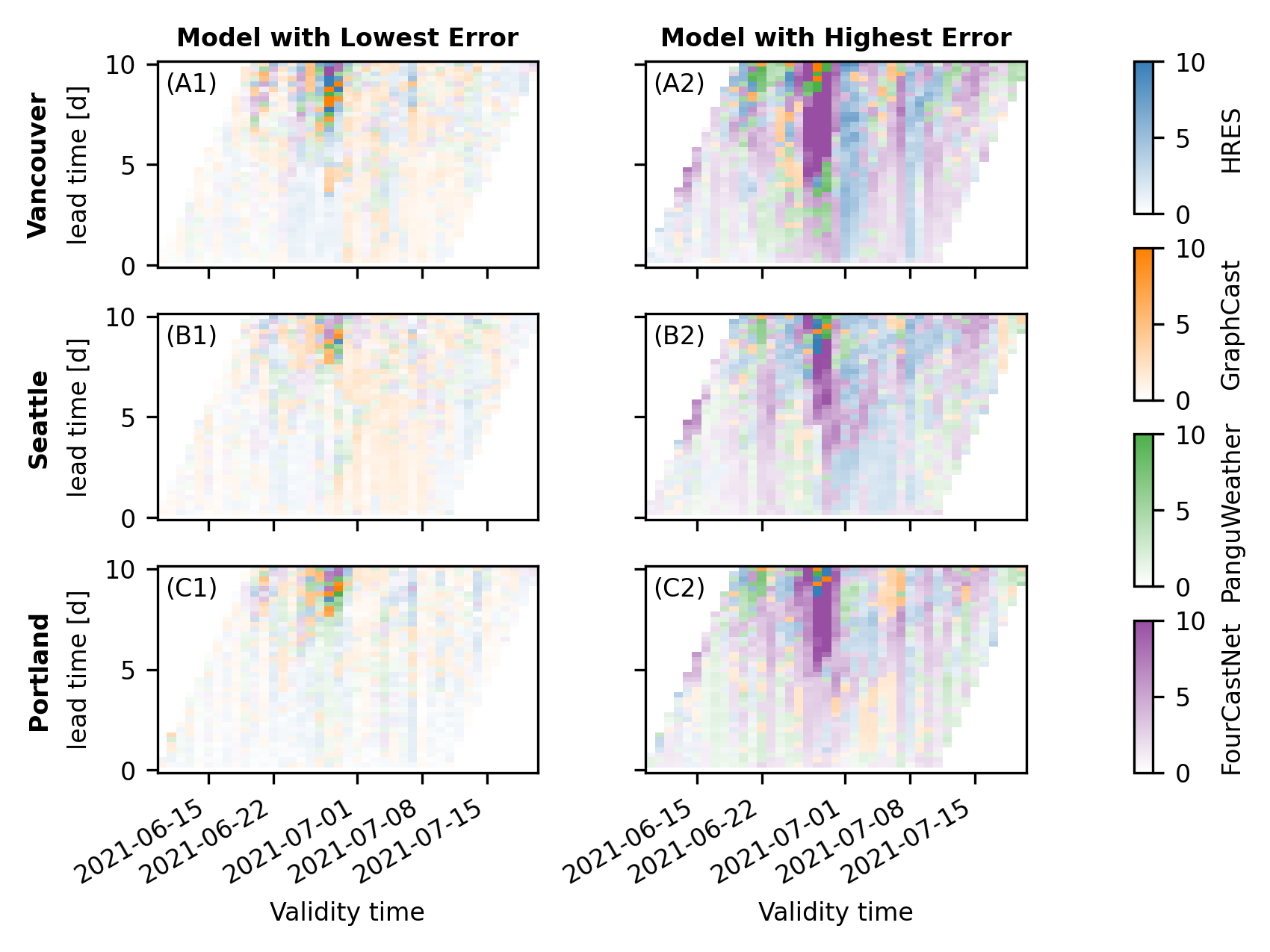}
    \caption[]{Predictability barrier plots after taking the argmin (left) and argmax (right) over the $\mathrm{RMSE}$ (in \unit{\kelvin}) of the different models. This means that the choice of colorbar in each pixel of the left panel indicates which model had the lowest $\mathrm{RMSE}$ for the given lead time and validity date, in the right panel it shows which model had the largest $\mathrm{RMSE}$.}
    \label{f:PNWBestWorst}
\end{figure*}

In this subsection, we show the additional \cref{f:PNW_RMSE2020_2022,f:PNWAnomalies,f:PNWBestWorst} for analysis of the 2021 Pacific Northwest Heatwave.

\clearpage
\section{Further Details and Analysis of the 2023 South Asian Humid Heatwave}
\label{As:AHH}
\subsection{Shape Files}
\label{ss:shape_files}
In \cref{ss:sa_humid_heatwave}, we use shape files to subset the study regions of the South Asian humid heatwave. We use country boundaries from the ``World Administrative Boundaries - Countries and Territories'' (Open Government License 3.0, \url{https://public.opendatasoft.com/explore/dataset/world-administrative-boundaries/information/}) data set by the World Food Programme and for the India-Bangladesh region we additionally use the (1976-2000) map of ``World Maps of the Köppen-Geiger Climate Classification''  (Creative Commons Attribution 4.0, \url{https://datacatalog.worldbank.org/search/dataset/0042325}). We only include grid cells with Köppen-Geiger Class A (``tropical'') in the India-Bangladesh region.

\subsection{Relative Humidity}
\label{ss:relative-humidity}

Relative humidity is required to compute the heat index, but PanguWeather \citep{biAccurateMediumrangeGlobal2023} and GraphCast \citep{lamLearningSkillfulMediumrange2023} only produce specific humidity, and only on atmospheric pressure levels, not at the surface. Therefore we first need to compute relative humidity from specific humidity. We exclude FourCastNet \citep{pathakFourCastNetGlobalDatadriven2022} from the case study in \cref{ss:sa_humid_heatwave} because here humidity is only available at pressure levels higher than \qty{850}{\hecto\pascal}. 

To convert specific humidity to relative humidity from the machine learning models, we first compute the saturation vapor pressure at the respective pressure level using the August–Roche–Magnus formula:
\begin{equation}
    \label{eq:august-magnus-roche}
    e_s(T) = \qty{6.1094}{\hecto\pascal} \exp\left(\frac{\num{17.625} T}{T + \num{243.04}}\right)
\end{equation}
where $T$ is the temperature at the respective pressure level in \unit{\celsius}. We compute the mixing ratio at saturation $r_s$ (dimensionless) from pressure and saturation vapor pressure:
\begin{equation}
    \label{eq:mixing-ratio}
    r_s = 0.622 \frac{e_s}{p - e_s}
\end{equation}
where $p$ is the pressure defining the pressure level. We then compute relative humidity $RH$ from the mixing ratio at saturation and the specific humidity $q$:
\begin{equation}
    \label{eq:relative-humidity-specific_humidity}
    RH = \frac{q}{(1-q) r_s},
\end{equation}
where specific humidity is given in \unit{\gram\per\kilo\gram}, and relative humidity is given in percent throughout the rest of the study. We emphasize again that $RH$ is not the relative humidity at the surface, which is not available for the ML models, but rather the humidity at pressure levels. 

For HRES and ERA5, it is additionally possible to compute relative humidity from $2m$ temperature $T_{2m}$ and $2m$ dewpoint temperature $T_d$:
\begin{equation}
    \label{eq:relative-humidity-dewpoint_temperature}
    RH = \frac{e_s(T_d)}{e_s(T_{2m})},
\end{equation}
where $e_s$ was computed using \cref{eq:august-magnus-roche} with $T_{2m}$ and $T_d$ given in degrees Celsius.

\subsection{Heat Index}
\label{ss:heat-index}
As described in \cref{ss:sa_humid_heatwave}, we follow \citet{zachariahExtremeHumidHeat2023a} during the computation of the heat index in using a modified version of the heat index \citep{rothfuszHeatIndexEquation1990} used by NOAA Weather Prediction Center (WPC). The NOAA WPC formulation is accessible at \url{https://www.wpc.ncep.noaa.gov/html/heatindex\_equation.shtml} (last accessed April 5 2024, page last modified May 12 2022 19:37:55 UTC). 

In the following, the heat index formula is given for temperature in \unit{\degree F}. We transform the temperature input to \unit{\degree F} and the heat index output to \unit{\celsius} in the study. 

To obtain the heat index $HI$, a simplified version $HI_s$ is computed first: 
\begin{equation}
    \label{eq:heat-index-simple}
    HI_s = 0.5 (T_{2m} + 61 + 1.2(T_{2m}-68) + 0.094RH)
\end{equation}
where $T_{2m}$ is the $2m$ temperature in \unit{\degree F}, and $RH$ is the relative humidity given as a percent value between \num{0} and \num{100}. If $(HI_s+T_{2m})/2$ is smaller than or equal to \qty{80}{\degree F}, the computation is complete, and the heat index is used as computed in \cref{eq:heat-index-simple}. Otherwise, the heat index is recomputed using a more elaborate formula:
\begin{equation}
\begin{split}
    \label{eq:heat-index-main}
    HI = &-42.379 + 2.04901523T_{2m} + 10.14333127 RH - 0.22475541 T_{2m} RH \\ &- 0.00683783 T_{2m}^2 - 0.05481717 RH^2 + 0.00122874 T_{2m}^2 RH \\ &+ 0.00085282 T_{2m} RH^2 - 0.00000199 T_{2m}^2 RH^2
\end{split}
\end{equation}

If $RH < \qty{13}{\percent}$ and $\qty{80}{\degree F} < T_{2m} < \qty{112}{\degree F}$, the heat index computed according to \cref{eq:heat-index-main} is corrected:

\begin{equation}
    \label{eq:heat-index-correction-1}
    HI = HI - \frac{13-RH}{4} \sqrt{(17-|T_{2m}-95|)/17}
\end{equation}

A different correction is applied if $RH > \qty{85}{\percent}$ and $\qty{80}{\degree F} < T_{2m} < \qty{87}{\degree F}$:
\begin{equation}
    \label{eq:heat-index-correction-2}
    HI = HI + \frac{(RH-85)}{10} \frac{87-T_{2m}}{5}
\end{equation}

We always use the \qty{2}{\meter} temperature $T_{2m}$ in the computation of the heat index. If the relative humidity is not available at the surface level, we specify in the text, how we substitute the value at the surface.

$HI$ values are classified as in \citet{zachariahExtremeHumidHeat2023a}, the classes are listed with potential health consequences in \cref{tab:hi_categories}.

\begin{table*}[tb]
\centering
\caption{Heat index categories, adapted from \citep{blazejczyk2012comparison}}
\begin{tabular}{p{3cm}p{35mm}p{85mm}}\toprule
Category & Definition & Possible heat disorders for people in high-risk groups \\\midrule
Low risk & $HI < \qty{27}{\celsius}$ & Fatigue possible with prolonged exposure and/or physical activity \\
Caution & $\qty{27}{\celsius} \leq HI < \qty{32}{\celsius}$ & Sunstroke, muscle cramps, and/or heat exhaustion possible with prolonged exposure and/or physical activity \\
Extreme caution & $\qty{32}{\celsius} \leq HI < \qty{41}{\celsius}$ & Sunstroke, muscle cramps, and/or heat exhaustion possible with prolonged exposure and/or physical activity \\
Danger & $\qty{41}{\celsius} \leq HI < \qty{54}{\celsius}$ & Sunstroke, muscle cramps, and/or heat exhaustion likely. Heatstroke possible with prolonged exposure and/or physical activity \\
Extreme danger & $\qty{54}{\celsius} \leq HI$ & Heat stroke or sunstroke likely \\
\bottomrule
\end{tabular}
\label{tab:hi_categories}
\end{table*}

\subsection{Additional Figures}
\begin{figure*}[tb]
    \centering
    \includegraphics[width=0.6\textwidth]{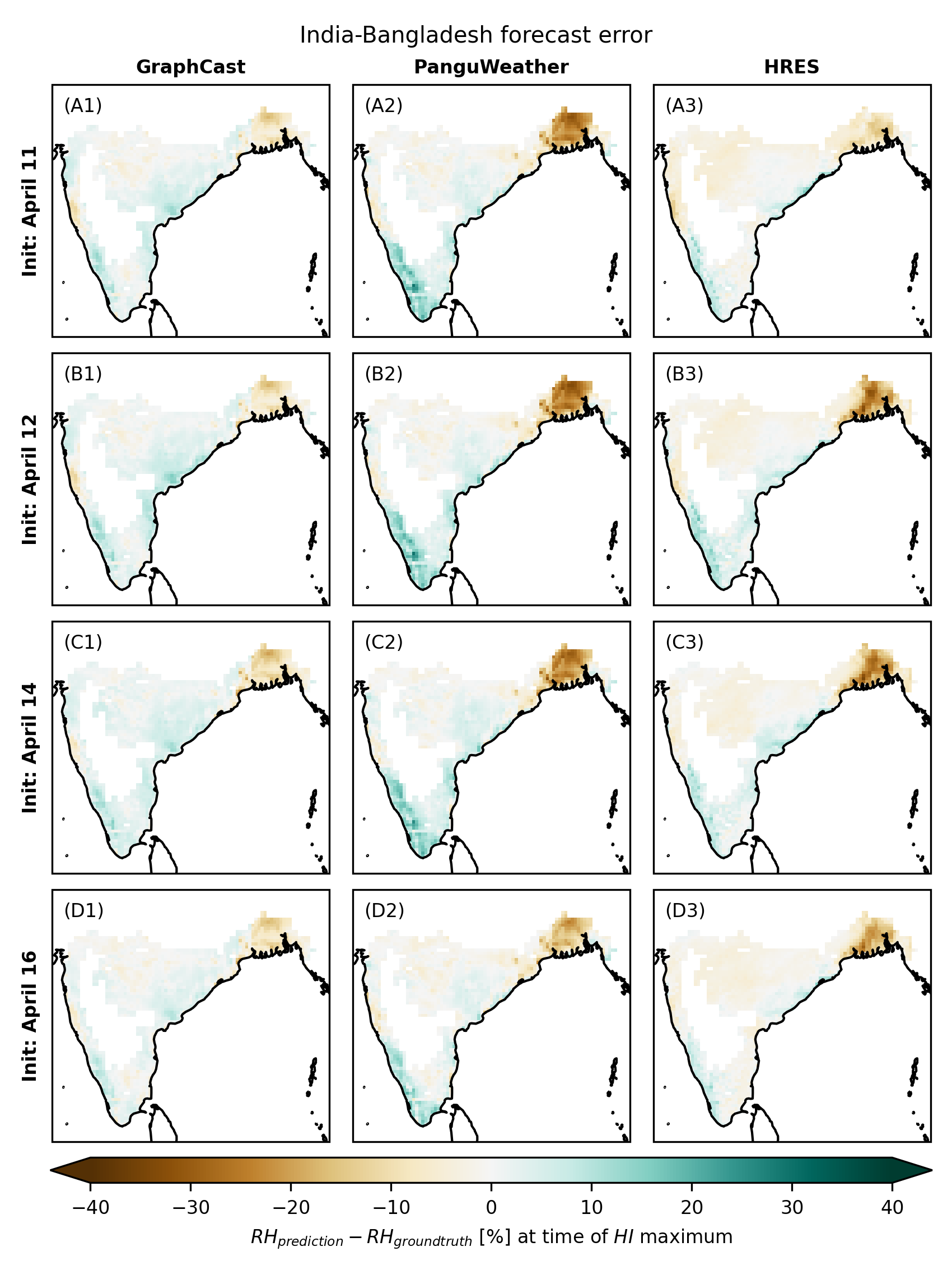}
    \caption[]{Forecast error for $RH_{\qty{1000}{\hecto\pascal}}$ at the time step of each day when observed $HI$ peaked in the corresponding ground truth data set, averaged over April~17--April~20.}
    \label{f:AHHspatialHumidityPressureLevel}
\end{figure*}

\begin{figure*}[tbp]
    \centering
    \includegraphics[width=0.6\textwidth]{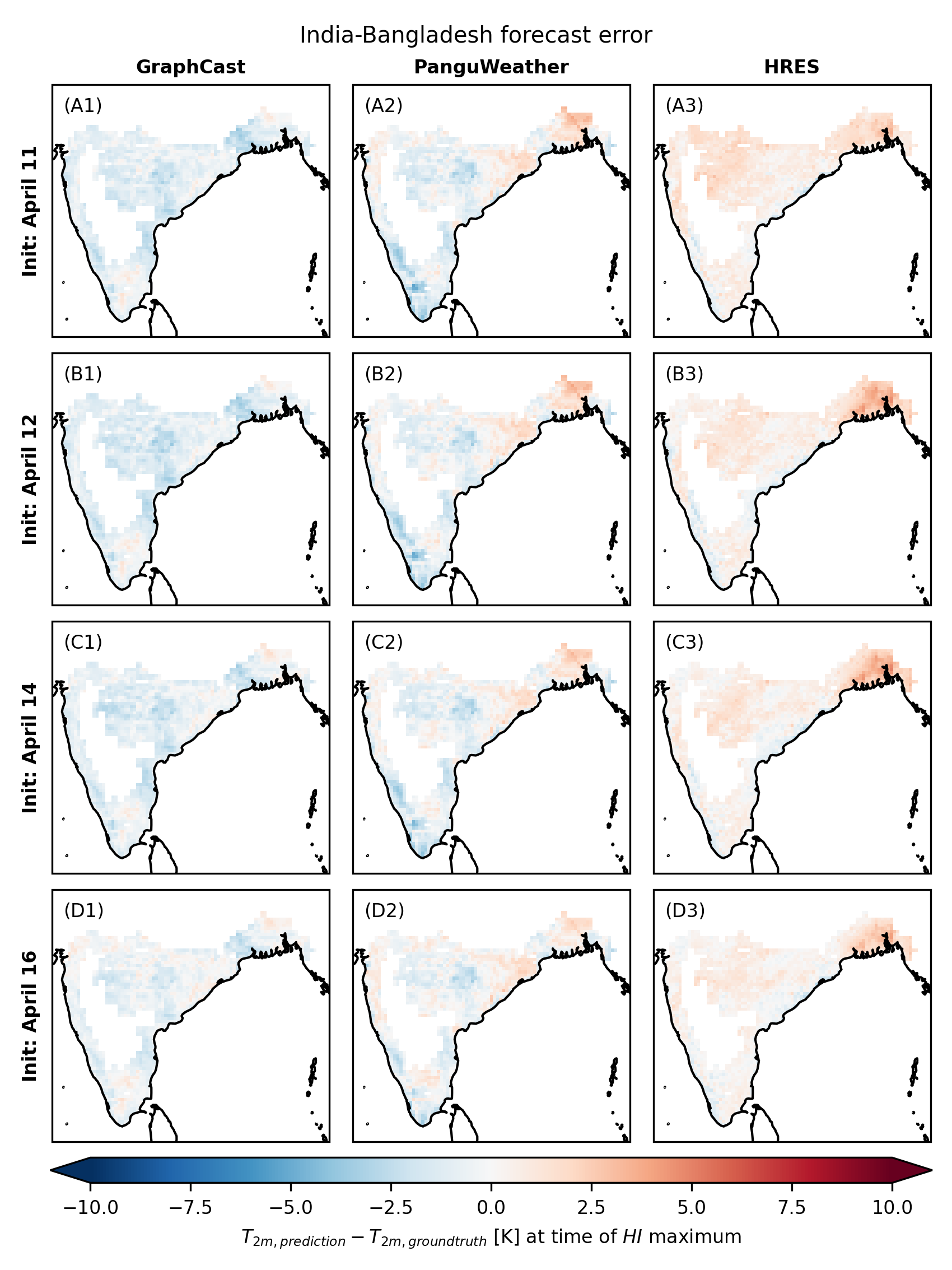}
    \caption[]{Forecast error for $T_{2m}$ at the time step of each day when observed $HI$ peaked in the corresponding ground truth data set, averaged over April~17--April~20.}
    \label{f:AHHspatialTemperaturePressureLevel}
\end{figure*}

\begin{figure*}[tbp]
    \centering
    \includegraphics[width=0.6\textwidth]{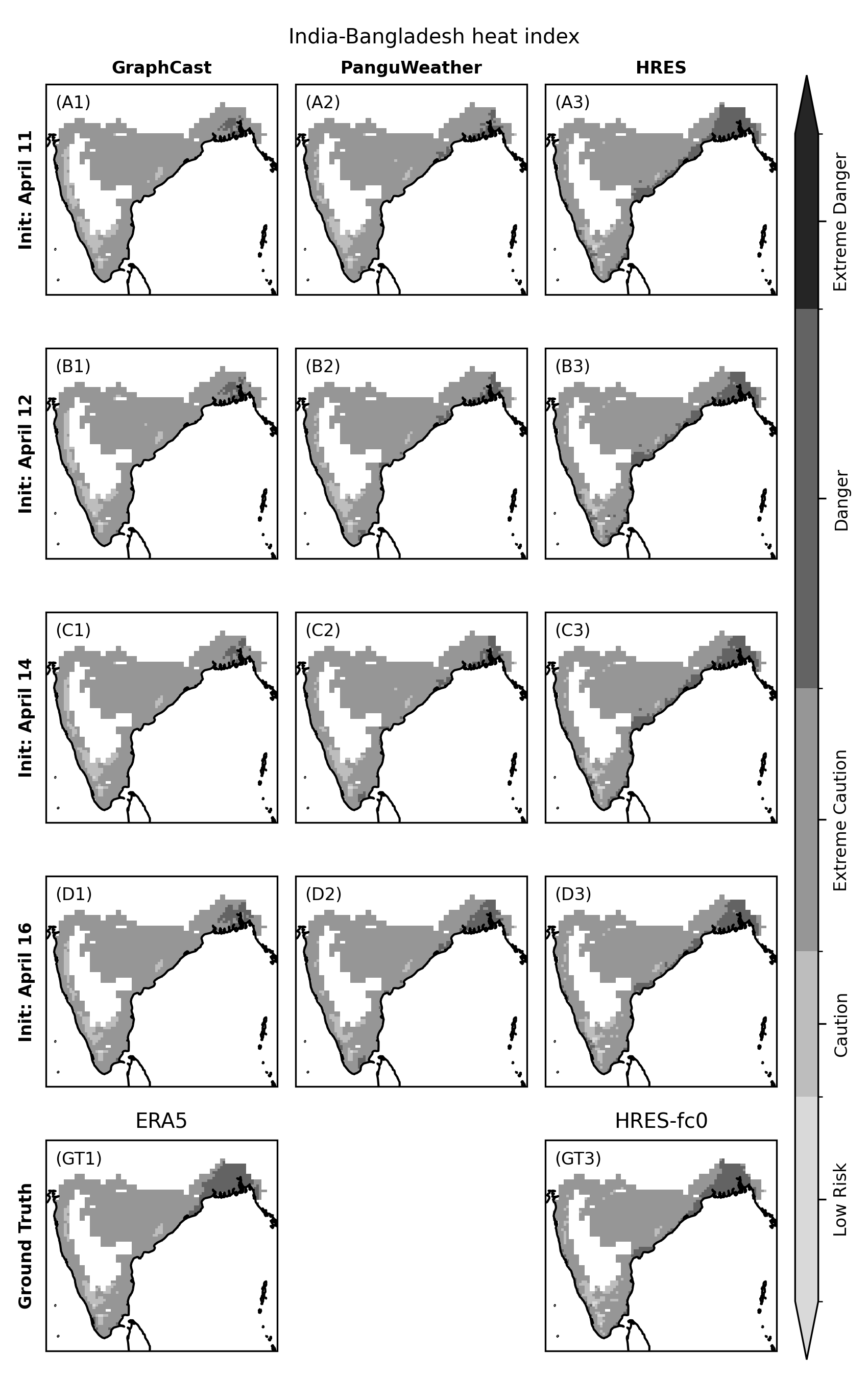}
    \caption[]{Category of mean daily maximum $HI$ (see \cref{tab:hi_categories}) predicted for April~17--April~20 with different models and for varying initialization times computed using $RH_{\qty{1000}{\hecto\pascal}}$. Forecasts are started at 00:00 UTC on the specified initial date. Panels (GT1), (GT3): categories in ground truth data sets.For ML models, $HI$ is computed using $RH_{\qty{1000}{\hecto\pascal}}$, while for all other panels we use $RH_{sfc}$.}
    \label{f:AHHcategoriesForecast}
\end{figure*}

In this subsection, we show the additional \cref{f:AHHspatialHumidityPressureLevel,f:AHHspatialTemperaturePressureLevel,f:AHHcategoriesForecast} for our analysis of the 2023 South Asian Humid Heatwave.

\clearpage
\section{Further Analysis of the 2021 North American Winter Storm}
\label{As:AWS}
\subsection{Additional Figures}
\begin{figure*}[bh]
    \centering
    \includegraphics[width=30pc]{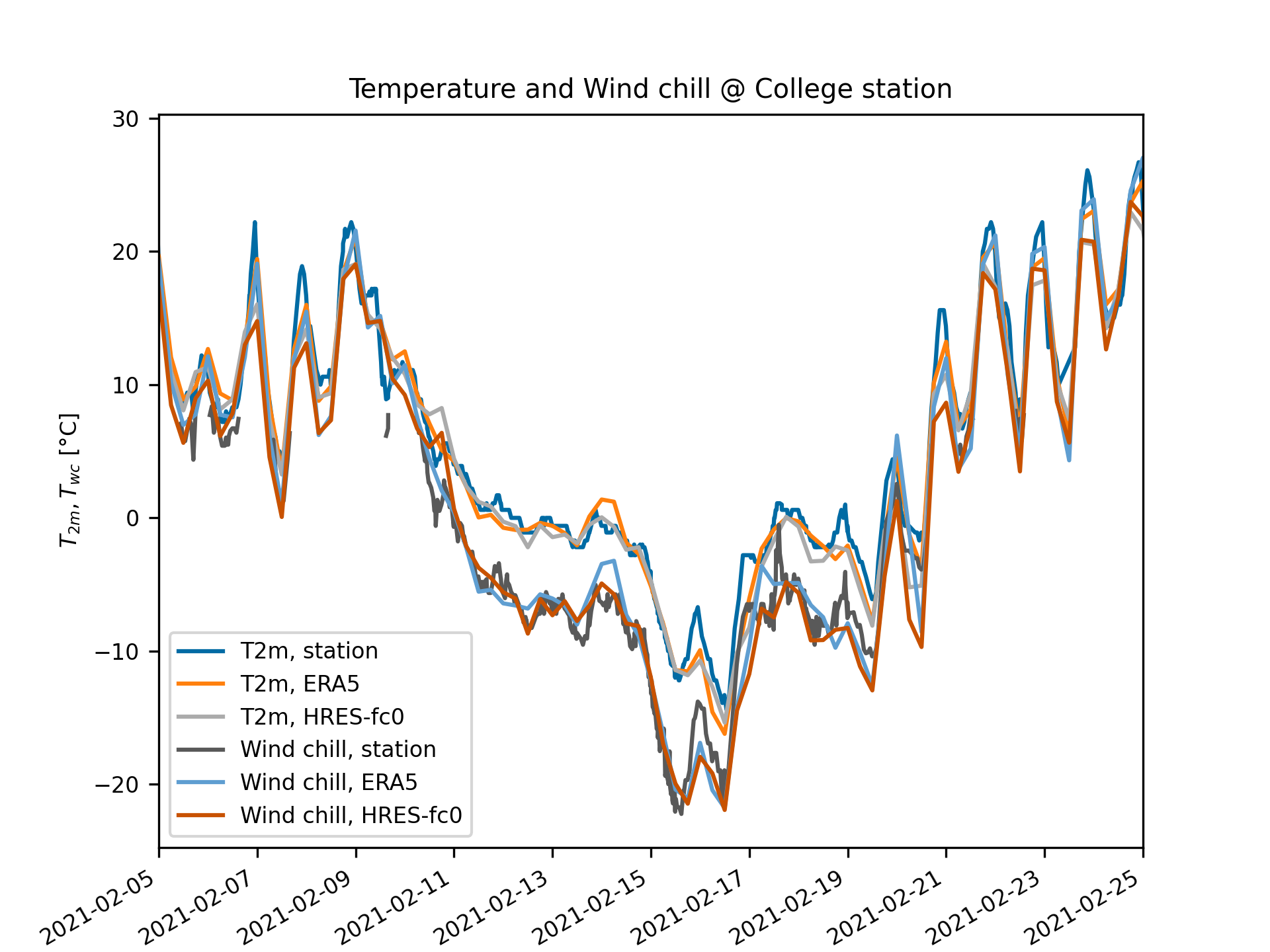}
    \caption[]{ERA5 and HRES-fc0 $T_{2m}$ and wind chill index $T_{wc}$ time series in the grid box closest to College Station, Texas. Weather station data: Easterwood Field, College Station, Texas. Data retrieved from Integrated Surface Dataset (ISD) \citep{smith2011integrated}. Figure ignores thresholds in the definition of $T_{wc}$ for better readability.}
    \label{f:AWSCollegeStation}
\end{figure*}

\begin{figure*}[tbp]
    \centering
    \includegraphics[width=27pc]{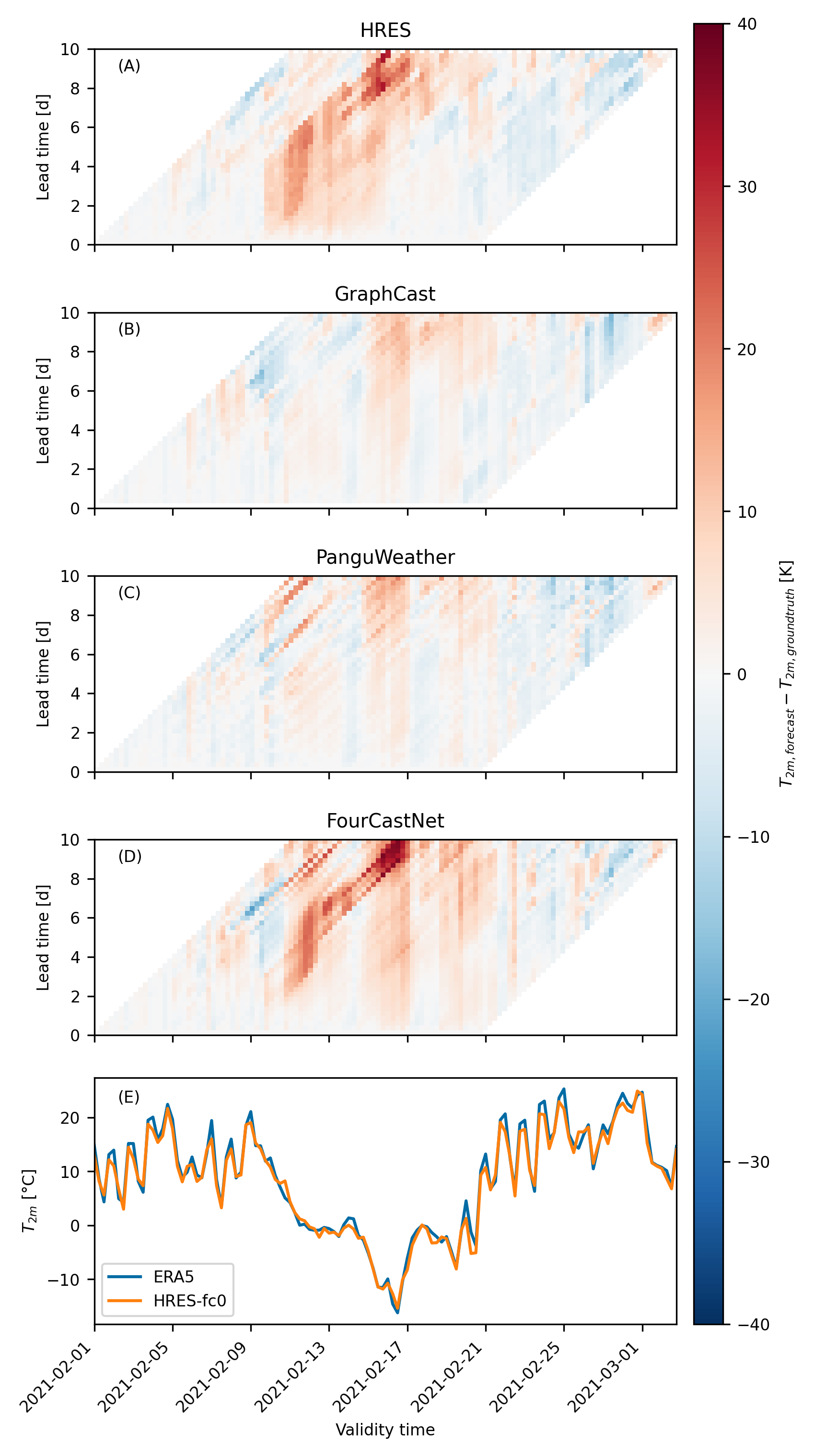}
    \caption[]{Panels (A) to (D): $T_{2m}$ prediction errors for different validity and lead times in the grid cell closest to College Station, Texas (UTC time used). Panel (E): ground truth $T_{2m}$ time series in the same grid cell.}
    \label{f:AWSPredBarrT2m}
\end{figure*}

\begin{figure*}[tbp]
    \centering
    \includegraphics[width=27pc]{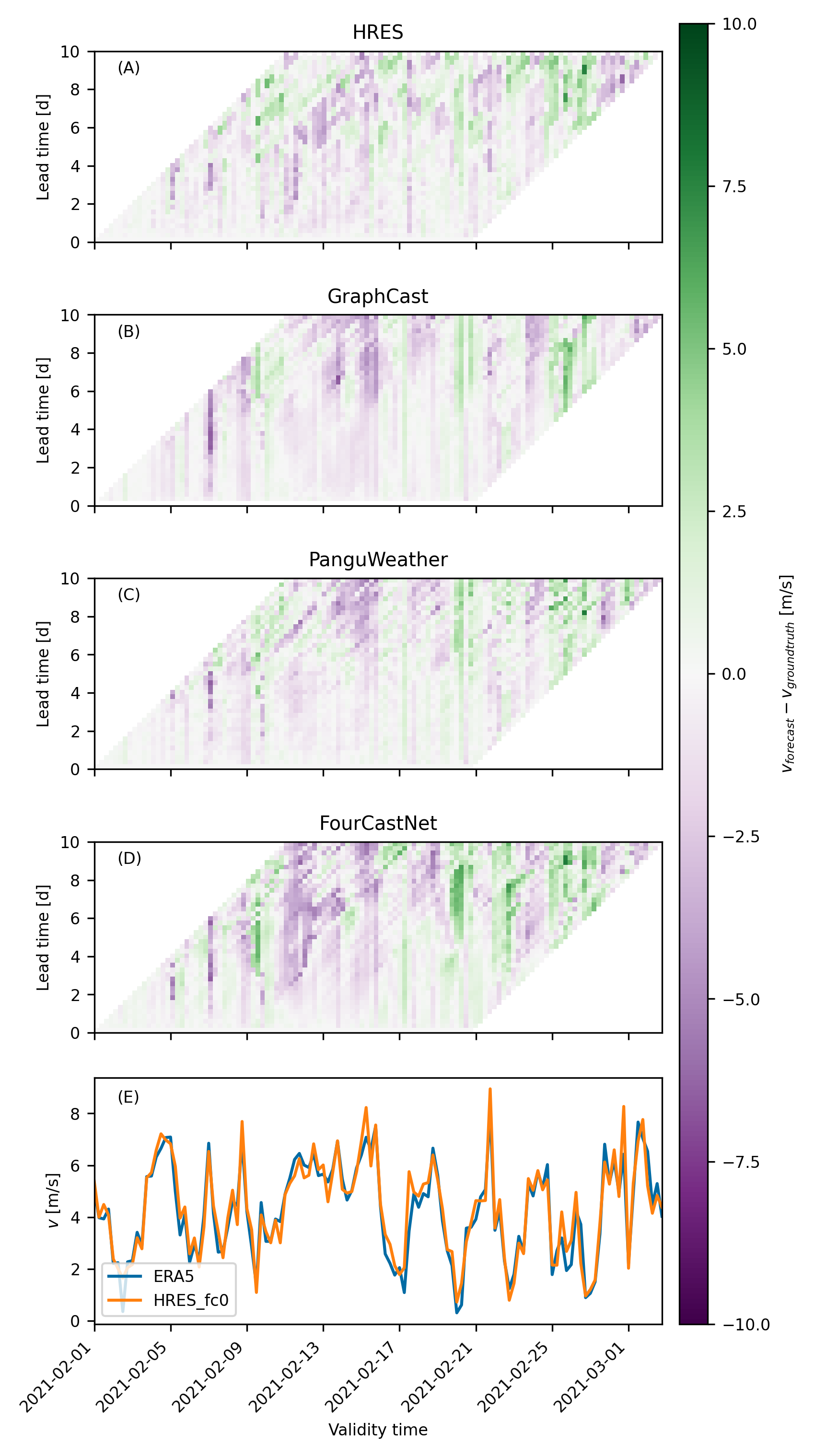}
    \caption[]{Panels (A) to (D): Prediction errors for surface wind speed for different validity and lead times in the grid cell closest to College Station, Texas (UTC time used). Panel (E): ground truth wind speed time series in the same grid cell}
    \label{f:AWSPredBarrWindspeed}
\end{figure*}

In this subsection, we show the additional \cref{f:AWSCollegeStation,f:AWSPredBarrT2m,f:AWSPredBarrWindspeed} for our analysis of the 2021 North American Winter Storm.

\vfill

\vfill

\clearpage

\newpage

\appendix

\section*{Declarations}

\paragraph*{Acknowledgements}
We thank Gloria Buriticá and Guohao Li for the discussions in early stages of the project, and Lily-Belle Sweet for her feedback on this manuscript. We are grateful to the developers of FourCastNet, PanguWeather, and GraphCast for sharing their code, and thank ECMWF for making their data sets publicly available, which has allowed us to conduct this study.

\paragraph*{Funding}
SE, OP, and ZZ acknowledge funding from the Swiss National Science Foundation Eccellenza grant ``Graph structures, sparsity and high-dimensional inference for extremes'' (grant no.~186858). 
JW acknowledges financial support by the Federal Ministry of Education and Research of Germany and by Sächsische Staatsministerium für Wissenschaft, Kultur und Tourismus in the program Center of Excellence for AI-research ``Center for Scalable Data Analytics and Artificial Intelligence Dresden/Leipzig'', project identification number: ScaDS.AI. JW and JZ acknowledge the Helmholtz Initiative and Networking Fund (Young Investigator Group COMPOUNDX, Grant Agreement VH-NG-1537).

\paragraph*{Author roles} 
Conceptualization: SE, OP, JW, ZZ, JZ; Data curation: OP, JW, ZZ; Formal analysis: OP, JW, ZZ; Funding acquisition: SE, JZ; Investigation: OP, JW, ZZ; Methodology: SE, OP, JW, ZZ, JZ; Project administration: SE, JZ; Resources: SE, JZ; Software: OP, JW, ZZ; Supervision: SE, JZ; Validation: OP, JW, ZZ; Visualization: OP, JW, ZZ; Writing - original draft: OP, JW, ZZ; Writing - review and editing: SE, OP, JW, ZZ, JZ. OP led the analysis and implementation of PanguWeather and FourCastNet, ZZ led the analysis and implementation of GraphCast, and JW led the case study data analysis. 

\paragraph*{Published article}
This document is the peer-reviewed ``Author's Accepted Manuscript'' of an article published in Artificial Intelligence for the Earth Systems~\citep{PascheDLWeatherExtremes}, with the DOI \href{https://doi.org/10.1175/AIES-D-24-0033.1}{\texttt{https://doi.org/10.1175/AIES-D-24-0033.1}}. 
When citing this work, please refer to the published version. 

\section*{Supplements}
\paragraph*{Supplementary materials}
The Supplementary Materials related to this paper are appended to this document.

\paragraph*{Data and code availability statement}
We use data from the ECMWF products ERA5, HRES, and TIGGE, which are published under a Creative Commons Attribution 4.0 International (CC BY 4.0) license. ERA5 is available on the Copernicus Climate Data Store. HRES forecasts initialized at 00:00/12:00 UTC can be accessed through ECMWF's TIGGE Data Retrieval portal. HRES forecasts initialized at 06:00/18:00 UTC are accessible through ECMWF's MARS, which requires access to be granted.
Recently, \citet{raspWeatherBenchBenchmarkNext2024} published cloud-optimized versions of the ERA5 and HRES data. We used these data sets for the case studies in 2021, and accessed their versions of ERA5, ERA5 climatology, and HRES forecasts initialized at 00:00/12:00 UTC.

Code to produce forecasts with GraphCast (\href{https://github.com/google-deepmind/graphcast}{\texttt{https://github.com/google-deepmind/\break{}graphcast}}), PanguWeather (\url{https://github.com/198808xc/Pangu-Weather}), and\break{} FourCastNet (\url{https://github.com/NVlabs/FourCastNet}) is available.

We published the preprocessed ground-truth data and model forecasts for the periods and regions studied \citep{PascheDLWFData2024}, under CC BY 4.0 licence. 
The code to reproduce the analyses and figures discussed in this work is available on \href{https://github.com/jonathanwider/DLWP-eval-extremes}{\texttt{https://github.com/jonathanwider/\break{}DLWP-eval-extremes}} (release v1.0).

\bibliographystyle{abbrvnat-namefirst}%
\begin{footnotesize}
\bibliography{ref_climate_transformers}
\end{footnotesize}

\clearpage

\appendix

\thispagestyle{plain} %

\vspace*{8mm}

\begin{center}
{\LARGE\bf 
Supplementary Materials to\\ 
``Validating Deep-Learning Weather Forecast Models on\\ 
Recent High-Impact Extreme Events''\\
}
\end{center}

\vspace{15mm}

\setcounter{section}{0}
\setcounter{subsection}{0}
\setcounter{equation}{0}
\setcounter{figure}{0}
\setcounter{table}{0}

\renewcommand{\thesection}{S.\arabic{section}}
\renewcommand{\thesubsection}{S.\arabic{section}.\arabic{subsection}}
\renewcommand{\theequation}{S.\arabic{equation}}
\renewcommand{\thefigure}{S.\arabic{figure}}
\renewcommand{\thetable}{S.\arabic{table}}

\renewcommand{\theHsection}{S.\thesection}
\renewcommand{\theHsubsection}{S.\thesubsection}
\renewcommand{\theHequation}{S.\theequation}
\renewcommand{\theHfigure}{S.\thefigure}
\renewcommand{\theHtable}{S.\thetable}

\section{Further Details on ML Models}
\label{sups:lists_variables}

In \cref{tab:ML_model_features_details} we list the surface and pressure-level variables predicted by the ML models.

\vfill

\begin{table*}[bh] %
\centering
\caption{Detailed list of variables predicted by ML models. $T_{2m}$: \qty{2}{m} temperature, $U_{10}, V_{10}$: horizontal components of \qty{10}{m} wind, $MSL$: mean sea-level pressure, $TP$: total precipitation, $SP$: surface pressure, $Z$: geopotential, $T$: Temperature, $RH$: relative humidity, $q$: specific humidity, $U,V$: horizontal components of wind, $W$: vertical component of wind}
\begin{tabular}{lccc}\toprule
   & FourCastNet & PanguWeather & GraphCast \\\midrule
Surface variables & $T_{2m}$ & $T_{2m}$ & $T_{2m}$ \\
& $U_{10}$ & $U_{10}$ & $U_{10}$ \\
& $V_{10}$ & $V_{10}$ & $V_{10}$ \\
& $MSL$ & $MSL$ & $MSL$ \\
& ($TP$) &  & $TP$ \\
& $SP$ &  &  \\\midrule
Atmospheric variables & $Z$ & $Z$ & $Z$ \\
 & $T$ & $T$ & $T$ \\
 & $RH$ & $q$ & $q$\\
 & $U$ & $U$ & $U$ \\
 & $V$ & $V$ & $V$ \\
 & & & $W$ \\\midrule
Levels [$\unit{\hecto\pascal}$]& 50, 500, 850, 1000 & 50, 100, 150, 200, & 1, 2, 3, 5, 7, 10, 20, 30, 50, \\
 & & 250, 300, 400, 500, 600, & 70, 100, 125, 150, 175, 200, \\
 & & 700, 850, 925, 1000 & 225, 250, 300, 350, 400, 450, \\
 & & & 500, 550, 600, 650, 700, 750, \\
 & & & 775, 800, 825, 850, 875, 900, \\
 & & & 925, 950, 975, 1000\\
\bottomrule
\end{tabular}
\label{tab:ML_model_features_details}
\end{table*}

\vfill
\newpage

\section{Further Analysis of the 2021 Pacific Northwest Heatwave}
\label{sups:PNW}

\begin{sidewaysfigure*}[htb]
    \centering
    \includegraphics[width=39pc]{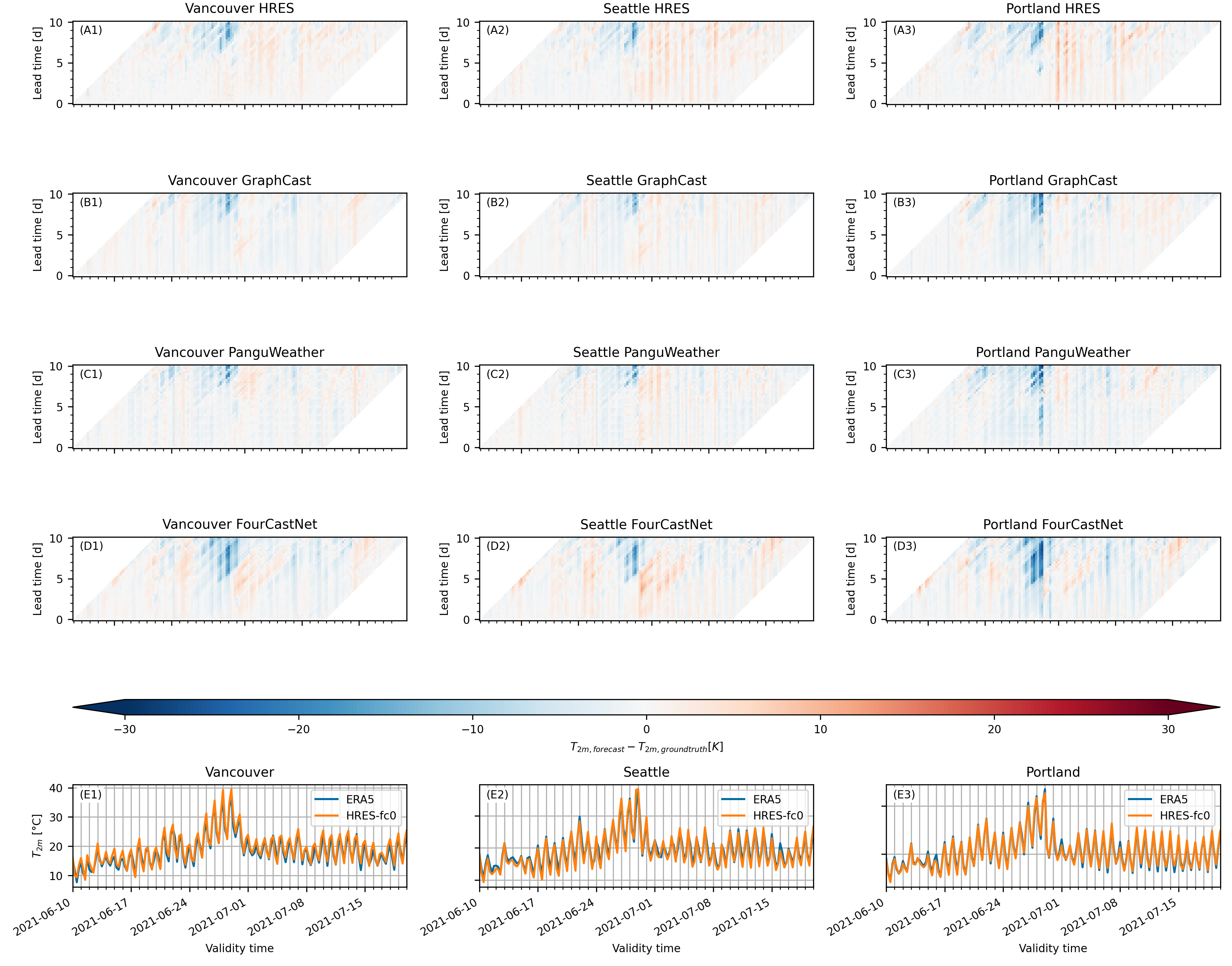}
    \caption[]{Panels (A1) to (D3): Barrier plots of forecast residuals for the grid cells closest to major cities affected by the 2021 heatwave. For HRES, HRES-fc0 is used as ground truth, for the ML models, we use ERA5 instead. Panels (E1) to (E3): time series of daily maximum $T_{2m}$ for the data sets used as ground truth.}
    \label{f:PNW_pred_all_alternative}
\end{sidewaysfigure*}

In \cref{f:PNW_pred_all_alternative}, we show an alternative version of Figure 2 in the main text. Here, we don't aggregate to daily scale and instead show the residuals $T_{2m, prediction} - T_{2m, ground truth}$. One can see patterns of the daily cycle in the predictions as well as negative forecast errors during the peak of the heatwave and positive forecast errors after the peak of the heatwave in some cases.

\begin{figure*}[tbp]
    \centering
    \includegraphics[width=39pc]{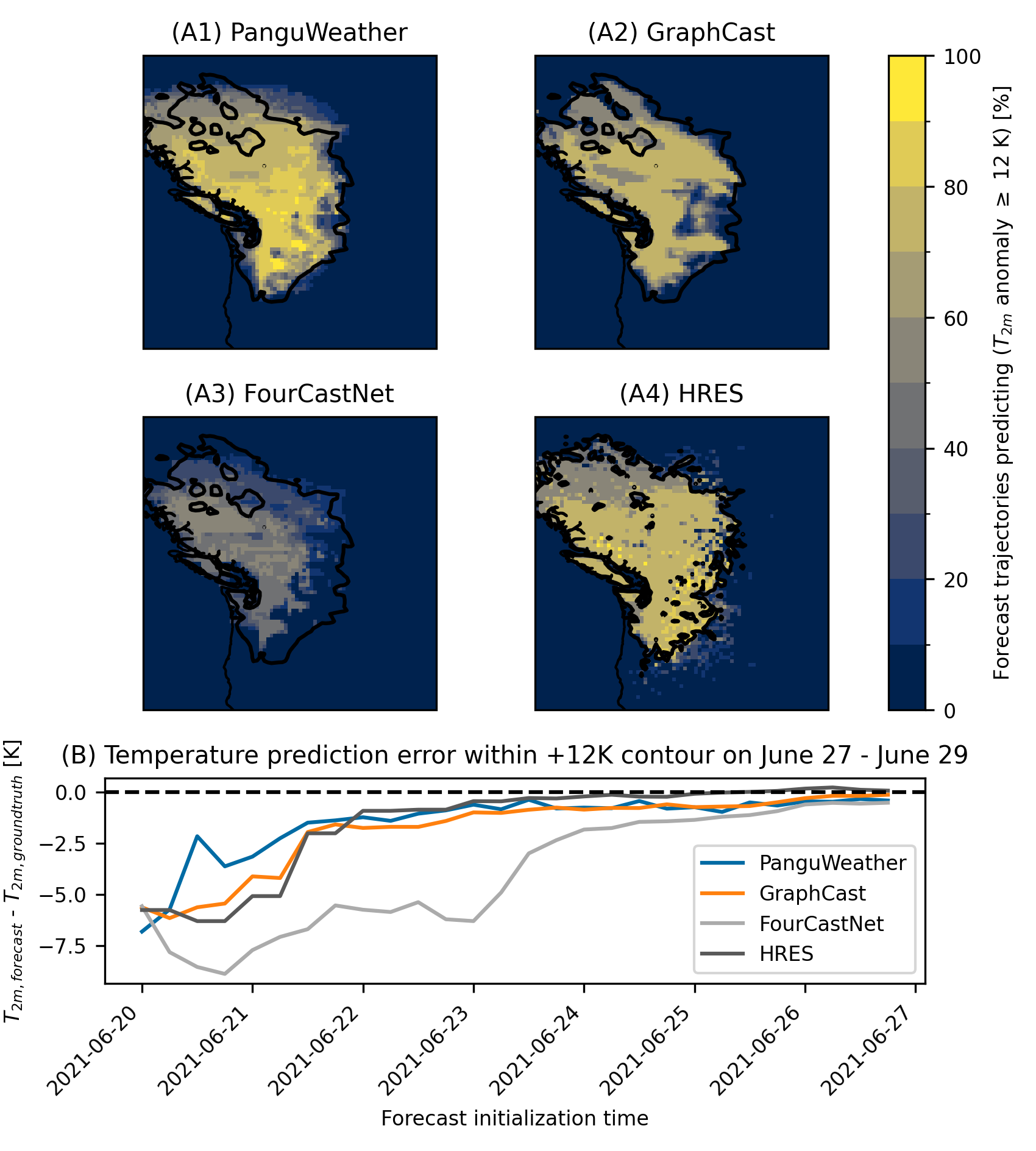}
    \caption[]{Panels (A1) to (A4): Black contour delimits the region in which the ground-truth average $T_{2m}$ anomaly between June 27 and June 29 exceeded \qty{12}{\kelvin}. The colormap indicates what fraction of forecasts started before the event predicted an anomaly exceeding \qty{12}{\kelvin} in the given grid box. Panel (B): error in the prediction of the June 27 to June 29 $T_{2m}$ anomaly, area-weighted average over the area within the \qty{12}{\kelvin} contours of panels (A1) to (A4).}
    \label{f:PNWSpatial}
\end{figure*}

To further analyze the spatial aspect of the predictions, we first compute $T_{2m}$ anomalies. We use the climatology provided by WeatherBench 2 \citep{raspWeatherBenchBenchmarkNext2024}, which is computed from ERA5 data between \qty{1990} and \qty{2019} (inclusive) and uses a sliding window of \qty{61}{\day} around the day of interest. For simplicity, we also use the ERA5 climatology in the computation of the HRES and HRES-fc0 anomalies, acknowledging that there are differences in the climatologies of ERA5 and HRES-fc0. In panels (A1) to (A4) of \cref{f:PNWSpatial}, we show in black the contour of the area in which the calculated true temperature anomaly, averaged between June 27 to 29 (inclusive), exceeds \qty{12}{\kelvin}.

We then look at forecasts initialized from \qty{7}{\day} to \qty{6}{\hour} (with a time step of \qty{6}{\hour}) before June 27, 00 UTC. For each grid cell, we determine the percentage of the forecasts that predict the average temperature anomaly to exceed \qty{12}{\kelvin} and plot this in panels (A1) to (A4) of \cref{f:PNWSpatial}. An ideal prediction system would always predict an anomaly of $\geq \qty{12}{\kelvin}$ everywhere within the black contour, and an anomaly $< \qty{12}{\kelvin}$ everywhere outside the contour.

Overall, the shapes of the predicted anomalies follow the shape of the true event well, i.e. heat was predicted in areas where it later actually occurred, and no large anomalies were predicted in regions where no large anomalies occurred. For FourCastNet, the anomaly is $< \qty{12}{\kelvin}$ everywhere for a large fraction of the forecasts - the yellow colors within the contour are faint. For GraphCast and HRES, the predicted area matches the contour well, meaning that a large fraction of the forecasts captured the right spatial distribution. For PanguWeather, more forecasts reach large positive anomalies than for GraphCast and HRES, this is reflected by brighter yellow colors. At the same time, PanguWeather is the only model whose forecasts ``spill out'' of the true \qty{12}{\kelvin} contour notably, i.e. for PanguWeather some of the forecasts predicted anomalies $\geq \qty{12}{\kelvin}$ in regions where they didn't occur in the ground truth data sets. 

Next, we compute the prediction error of the June 27 to 29 temperature anomaly for each grid cell and then take the (area-weighted) average over all pixels within the area in which the true anomaly exceeded \qty{12}{\kelvin} over these days. The results for the different models are shown in panel~(B) of \cref{f:PNWSpatial}. FourCastNet strongly under-predicts the temperature anomaly within the contour for initialization times before June 24, while the other models achieve better predictions of the magnitude of the heatwave sooner. PanguWeather's forecasts initialized on June 20 and June 21 predict notably warmer temperatures than HRES and GraphCast. PanguWeather and GraphCast slightly under-predict the size of the anomaly until the start of the event, while for HRES the under-prediction is smaller. One needs to keep in mind that the ERA5 ground truth and the HRES-fc0 ground truth don't coincide exactly, and thus it is not clear whether one can attribute this slight under-predictions of GraphCast and PanguWeather to model deficiencies.

\clearpage

\section{Further Analysis of the 2023 South Asian Humid Heatwave}
\label{sups:AHH}

\subsection{Humid Heatwave in the Laos-Thailand region}
\label{ss:laos-thailand}

In section 3.2 of the main paper, we analyzed how well the 2023 South Asia humid heatwave is predicted in the India-Bangladesh study region. Conclusions for the Laos-Thailand study region are similar. Here the analyzed heatwave peak is April~18--April~21 2023.

\begin{figure*}[hb] %
    \centering
    \includegraphics[width=27pc]{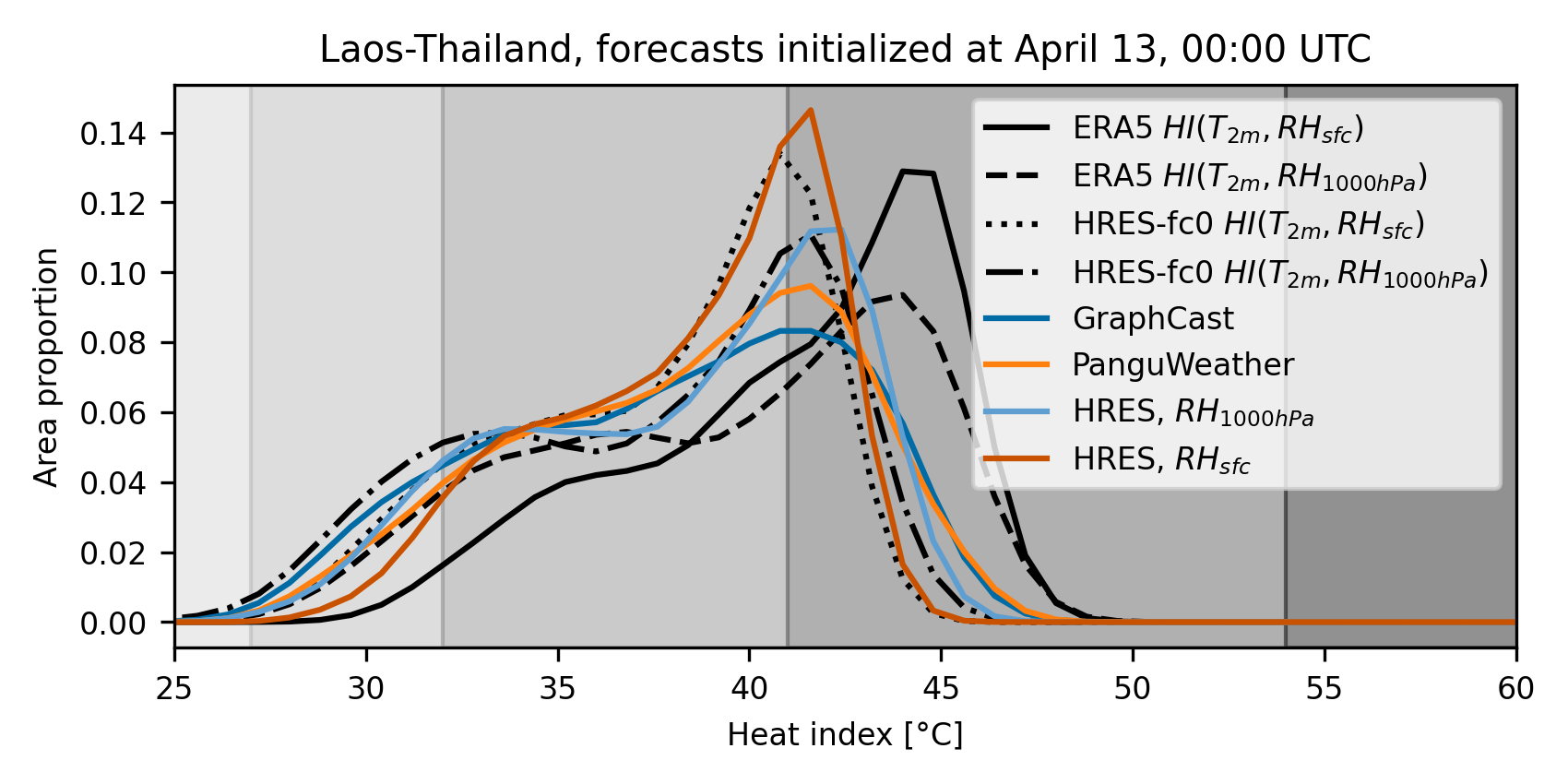}
    \caption[]{Proportion of area in study region with given mean daily maximum heat index during April~18––21, computed using area-weighted kernel density estimation. Shaded areas in the background indicate threat levels \citep{blazejczyk2012comparison}. From light gray to dark gray: low risk, caution, extreme caution, danger, extreme danger. Compared are distributions resulting from forecasts initialized 6  days prior to the start of the event and different ground truths: ERA5 and HRES-fc0, each in two versions of computing the heat index, either using $RH_{sfc}$ or using the substitute $RH_{\qty{1000}{\hecto\pascal}}$. For HRES forecasts, we show versions computed with $RH_{\qty{1000}{\hecto\pascal}}$ and $RH_{sfc}$ as well.}
    \label{f:AHHHIAreaLaosThailand}
\end{figure*}

In \cref{f:AHHHIAreaLaosThailand}, again the ground truth ERA5 and HRES-fc0 data sets don't yield the same $HI$ values. HRES follows its ground truth well, while the ML prediction models under-predict the high $HI$ values when compared to their ERA5 ground truth computed using $RH_{\qty{1000}{\hecto\pascal}}$ and the version computed with $RH_{sfc}$.

\begin{figure*}[tbp]
    \centering
    \includegraphics[width=27pc]{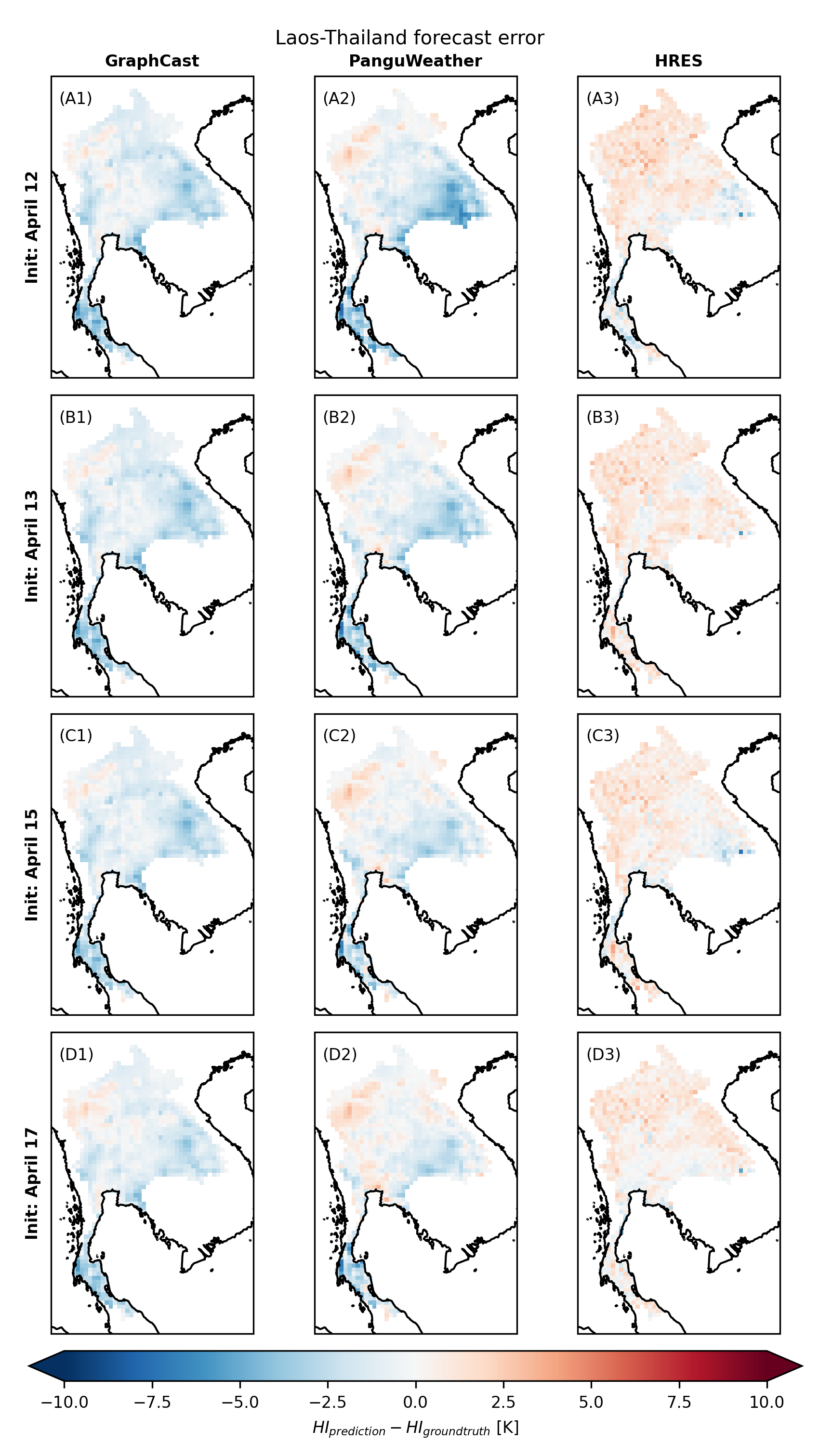}
    \caption[]{Error of the ${HI}$ prediction, for the time step of each day during which $HI$ peaked in the ground truth data set, averaged over April~18--21. For all forecasting methods and ground truth data sets, ${HI}$ is computed using ${RH}_{\qty{1000}{\hecto\pascal}}$ rather than the value at the surface.}
    \label{f:AHHspatialInitPressureLevelLaosThailand}
\end{figure*}

An under-prediction of $HI$ can also be seen in \cref{f:AHHspatialInitPressureLevelLaosThailand}, where the $HI$ forecast errors for the different models are compared. All computations for this plot use $RH_{\qty{1000}{\hecto\pascal}}$ rather than $RH_{sfc}$. There seems to be an under-prediction of $HI$ by the ML methods, while HRES values tend to be slightly larger than the HRES-fc0 ground truth.

\vfill

\vfill

\end{document}